\documentclass[preprint,12pt]{elsarticle}

\usepackage{graphicx}
\usepackage{amssymb}
\usepackage{textcomp}
\usepackage{url}
\usepackage{subfig}
\usepackage[top=1in, bottom=1in, left = 1in, right = 1in]{geometry}
\usepackage[utf8]{inputenc}
\usepackage{nicefrac}
\usepackage{multirow}
\usepackage[export]{adjustbox}
\usepackage{color, soul}
\usepackage{siunitx}

\biboptions{square, sort&compress}

\journal{Computational Materials Science}

\begin{document}

\begin{frontmatter}

\title{Phase Field Benchmark Problems for Dendritic Growth and Linear Elasticity}

\author[CHiMaD,ANL-2]{A. M. Jokisaari\corref{cor1}}
\ead{andrea.jokisaari@inl.gov}
\cortext[cor1]{Corresponding author, current affiliation Nuclear Science and Technology Directorate, Idaho National Laboratory, P.O. Box 1625, MS 3840, Idaho Falls, ID 83415}

\author[CHiMaD,NU]{P. W. Voorhees}
\ead{p-voorhees@northwestern.edu}

\author[NIST]{J. E. Guyer}
\ead{jonathan.guyer@nist.gov}

\author[NIST]{J. A. Warren}
\ead{james.warren@nist.gov}

\author[NAISE,ANL-1]{O. G. Heinonen}
\ead{heinonen@anl.gov}

\address[CHiMaD]{Center for Hierarchical Materials Design, Northwestern University, 2205 Tech Drive, Suite 1160, Evanston, IL, 60208, USA}
\address[ANL-2]{Materials Science Division, Argonne National Laboratory, 9700 South Cass Avenue, Lemont, IL 60439, USA}
\address[NU]{Department of Materials Science and Engineering, Northwestern University, 2220 Campus Drive, Evanston, IL 60208, USA}
\address[NIST]{Material Measurement Laboratory, National Institute of Standards and Technology, 100 Bureau Drive, MS 8300, Gaithersburg, MD 20899-8300}
\address[NAISE]{Northwestern-Argonne Institute of Science and Engineering, 2205 Tech Drive, Suite 1160, Evanston, Illinois 60208, USA}
\address[ANL-1]{Materials Science Division, Argonne National Laboratory, 9700 South Cass Avenue, Lemont, IL 60439, USA}

\begin{abstract}

We present the second set of benchmark problems for phase field models that are being jointly developed by the Center for Hierarchical Materials Design (CHiMaD) and the National Institute of Standards and Technology (NIST) along with input from other members in the phase field community.  As the integrated computational materials engineering (ICME) approach to materials design has gained traction, there is an increasing need for quantitative phase field results.  New algorithms and numerical implementations increase computational capabilities, necessitating standard problems to evaluate their impact on simulated microstructure evolution as well as their computational performance.  We propose one benchmark problem for solidification and dendritic growth in a single-component system, and one problem for linear elasticity via the shape evolution of an elastically constrained precipitate.  We demonstrate the utility and sensitivity of the benchmark problems by comparing the results of 1) dendritic growth simulations performed with different time integrators and 2) elastically constrained precipitate simulations with different precipitate sizes, initial conditions, and elastic moduli.  These numerical benchmark problems will provide a consistent basis for evaluating different algorithms, both existing and those to be developed in the future, for accuracy and computational efficiency when applied to simulate physics often incorporated in phase field models.

\end{abstract}


\end{frontmatter}

\section{Introduction \label{sec:Introduction}}

Over the past decade, the concept of integrated computational materials engineering (ICME) has firmly taken root within the materials science and engineering community. In the ICME approach, a material is described and modeled at different fundamental length and time scales; information is linked across these scales to fully capture material behavior and to develop the processing-structure-properties relationships used by materials engineers  \cite{allison2006integrated}.  The ICME approach may be applied to achieve different goals, such as the development of wholly new materials for specific applications or the tuning of an existing material's composition to meet multiple, unrelated (e.g., color and strength) design criteria.  The ICME approach has engendered multiple successes thus far, including the development of new alloys for aerospace \cite{kuehmann2008ultra,grabowski2013integrated}, automotive \cite{joost2012reducing, taub2015advanced}, and coinage \cite{Lass2015Alloy} applications.  

The phase field approach is one modeling technique that is commonly included in ICME frameworks.  Phase field modeling is a continuum method that is applied to study phenomena occurring on diffusive length and time scales (nanometers to micrometers and microseconds or longer). In a phase field model, certain field variables are defined (e.g., solute concentration) that are continuous across the computational domain and that vary smoothly across phase interfaces.  A free energy of the system is defined based on these field variables, and the evolution of the system is driven by a reduction of the free energy.  Unlike mean-field models, which average out spatial variations in microstructure, phase field models resolve microstructure evolution in space, elucidating how variations within a microstructure form and interact.  Historically, phase field models have tended to be more qualitative in nature, providing insights into potential microstructure evolution mechanisms.  However, with the integration of realistic energetics (e.g., CALPHAD-based free energies \cite{grafe2000coupling}), phase field models are now being crafted to quantitatively describe real materials. They have been applied to study a range of phenomena, such as the ``rhenium effect'' in nickel-based superalloys \cite{mushongera2015effect} and the formation of gas bubble superlattices in uranium-molybdenum nuclear fuel \cite{hu2016formation}.  For comprehensive descriptions and reviews of phase field modeling, see Refs.~\cite{ASM,boettinger2002phase,chen2002phase,emmerich2008advances,moelans2008introduction,steinbach2009phase,nestler2011phase,steinbach2013phase}.

Error and uncertainty in models and simulation results are important considerations when applying the ICME approach to materials design \cite{TMS2015Modeling}.  Error, broadly speaking, is the deviation in modeled results from physical reality, while uncertainty is the likelihood of the model and implementation to reproduce the physical phenomenon in question with some variation in parameters. Some degree of error and uncertainty exist within any model results, and the linked nature of ICME frameworks means that error and uncertainty at one level may propagate through the results at different length and time scales. While error analyses for phase field models and algorithms are published on occasion (e.g., Refs.~\cite{karma1998quantitative,zhang2013quantitative,jokisaari2016nucleation}), the proliferation of scientific publications can make it difficult for new modelers to fully familiarize themselves with existing analyses on the subject, which are often a part of a larger publication on a different topic.  The development of new solver frameworks combined with the growing popularity of the phase field approach and a desire for ``turnkey'' simulation software means that a common basis is needed for performing error analyses and uncertainty quantification; new numerical methods must be evaluated for their accuracy with respect to phase field model physics. 

To help address these issues, the Center for Hierarchical Materials Design (CHiMaD) and the National Institute of Standards and Technology (NIST) are developing a suite of numerical benchmark problems that will allow the uniform testing of phase field algorithms and implementations.  We choose problem formulations and benchmark metrics that balance the need for nontrivial solutions with computational resource requirements and the investment of human effort, and we intend that these problems are used by the community to understand how numerical methods impact results by comparing different results  (see Ref.~\cite{jokisaari2017benchmark} for a detailed discussion). Importantly, the benchmark problem effort includes significant community involvement regarding problem proposal, design and vetting. These numerical benchmark problems are a necessary precursor before validating a model to experimental results, as the correctness of the numerical method must be verified for validation studies to be useful.  Each of these benchmark problems targets a specific aspect of physics commonly found in phase field models. In particular, we address models that include coupled physics, as these can pose significant numerical challenges.  The first set of problems focus on the diffusion of solute and the growth of a second phase \cite{jokisaari2017benchmark}, while this set involves solidification and dendritic growth as well as linear elasticity.  To support community involvement, we have developed a website (\url{https://pages.nist.gov/pfhub/}), which serves as a repository for the problem statements and the results from different numerical implementations.  

We choose dendritic growth as a subject both because of the importance of the physical phenomenon in controlling materials properties \cite{ASM2004metals} and because of the sensitivity of simulation results on phase field model formulations and implementations (see, for example, Refs.~\cite{karma1998quantitative,warren1995prediction}). Historically, dendritic growth is one of the first applications of phase field modeling \cite{Fix,Langer}, and remains a significant area of research today.  Sharp interface limit \cite{Langer,collins1985diffuse,kobayashi1993modeling,mcfadden1993phase,wang1993thermodynamically} and thin interface limit \cite{karma1996phase,karma1998quantitative,mcfadden2000thin,karma2001phase} analyses show that the diffuse-interface phase field formulation is asymptotically equivalent to the sharp-interface Stefan formulation.  With the introduction of an ``anti-trapping current'' to correct for solute trapping due to the jump in chemical potential at the solid/liquid interface \cite{karma2001phase,echebarria2004quantitative}, quantitative phase field modeling of alloy solidification can be performed using unphysically large diffuse interface widths.  Today, massive increases in computing power and the advent of scientific computing on graphical processing units (GPUs) enable large-scale, quantitative 3D phase field simulations (see, for example, Refs.~\cite{hotzer2015large,shibuta2015solidification} and reviews \cite{takaki2014phase,granasy2014phase}).

Similar considerations drive our choice for the selection of linear elasticity as the subject of the second benchmark problem.  Like dendrites, precipitates are a key microstructural feature impacting the strength of alloys \cite{chawla1999mechanical}, and they are often elastically stressed, which affects their shape and their microstructure evolution during service.  Elasticity has long been incorporated into phase field models: indeed, Cahn's seminal paper on spinodal decomposition \cite{cahn1961spinodal} incorporates elastic strains due to composition fluctuations.  Eshelby presents an analytical solution for the elastic field of a single coherent, elastically stressed precipitate in an infinite matrix \cite{eshelby1957determination}, but the generalized problem of multiple interacting precipitates in a matrix with arbitrary crystal structure, lattice parameter misfit and elastic stiffnesses can only be solved numerically.  Sharp-interface approaches provide insight into equilibrium elastic shapes and coarsening under the influence of elastic stress \cite{voorhees1992morphological,thompson1994equilibrium,su1996dynamicsI,su1996dynamicsII,akaiwa2001large}, but these approaches have difficulty simulating precipitate splitting or merging. Early phase field formulations studying elastically stressed precipitates demonstrate the power of the method (e.g., Refs.~\cite{wang1993shape,wang1994effect,wang1995shape}), and present-day studies have expanded to 3D simulations (e.g., Refs.~\cite{goerler2017topological,radhakrishnan2016phase,shi2015microstructure,cottura2015role}) and formulations that include plasticity (e.g., Refs.~\cite{cottura2016coupling,ammar2014modelling,guo2008elastoplastic}).
 
In this work, we present the second set of community-driven benchmark problems developed by CHiMaD and NIST.  One problem targets solidification by modeling dendritic growth for a single-component system, and the other  targets linear elasticity by simulating the equilibrium shape of an elastically stressed precipitate. We discuss the importance of these problems for evaluating new numerical algorithms, a growing concern given the rise of generalized numerical solver frameworks that may include new algorithms that may not be suitable for phase field model physics.  We demonstrate the utility of the dendritic growth problem by performing the same simulations with different time integration algorithms.  In addition, we discuss how the problems may be modified to test small variations in model formulation or parameterization, and show how the shape of an elastically constrained precipitate is affected by these variations.  Finally, we urge researchers to provide feedback on the existing benchmark problem set, contribute their results to the website, and make suggestions for modifications or future benchmark problem topics.

\section{Benchmark problem formulations \label{sec:Model-formulation}}

The phase field approach is well suited for modeling multiphysics problems, that is, phenomena driven by more than one physical factor, e.g., diffusion in an electric potential field.  In a phase field model, a microstructure is defined by one or more field variables, or ``order parameters,'' which exist over the entire computational domain and which evolve in time.  Multiphysics models may define additional field variables, such as temperature, that are needed to describe the system but not the microstructure itself. The total system free energy, $\mathcal{F}$, is a functional of different local free energy density contributions that depend on the field variables.  These different free energy densities capture different energetic contributions to the system, such as bulk chemical energy, interfacial energy, elastic energy, etc. The time evolution of the order parameters is governed by functional derivatives of the free energy as driving forces (Onsager non-equilibrium thermodynamics), though additional field variables may be driven by different dynamics (e.g., thermal diffusion). If the relaxation dynamics of one field variable is orders of magnitude faster than another, the model may be formulated using a quasi-static approximation, that is the time-independent solution for the rapidly-evolving variable is computed at each time step of the slower-changing variable's evolution.

The two benchmark problem formulations presented in Section 2 involve multiphysics coupling: the model for solidification and dendritic growth incorporates anisotropic interfacial energy and the release of latent heat, and the model for the elastically constrained precipitate adds the physics of linear elastic solid mechanics to the Cahn-Hilliard equation.  We discuss each model formulation and parameterization, as well as our choices for computational domain sizes and initial and boundary conditions.  Both models are restricted to two dimensions to capture essential physics and potential computational pitfalls without requiring the use of significant computational resources.

\subsection{Solidification and dendritic growth \label{sub:solidification}}

Many commercial manufacturing processes, such as casting, welding, and additive manufacturing, result in the formation of dendrites during solidification.  The presence of a dendritic structure and the exact nature of that microstructure can have a significant impact on the mechanical properties of a workpiece, including the yield strength, ultimate strength, and ductility.  Modern phase field formulations of solidification and dendritic growth can be quite complex, involving multiple chemical components and fluid flow. However, we follow the benchmark problem design principles given in Ref.~\cite{jokisaari2017benchmark} to create (or select from the literature) a problem that captures the essential physics of the given phenomenon, yet is simplified and easy to implement. This is particularly important for the dendritic growth problem, as simulated microstructures can be extremely sensitive to numerical and model parameterization choices  \cite{warren1995prediction,karma1998quantitative}.  We specifically choose model parameters that will stress numerical solvers yet avoid the formation of instabilities, as simulation results would then be impossible to compare across implementations.

For this problem, we feel it is important to select an existing formulation from the literature, as a large body of work already exists on which to build community benchmarking efforts.  We select the isothermal variational formulation (IVF) of solidification for a single-component, undercooled melt system that is analyzed in Ref.~\cite{karma1998quantitative}.  This choice has the added benefit that results are tabulated for the corresponding sharp-interface model \cite{karma1998quantitative}, providing an additional basis for comparison.

\subsubsection{Model formulation \label{ssub:solidification-model}}

In this formulation, one order parameter, $\phi$, and one additional field variable, $U$, are evolved.  The phase of the material is described by $\phi$, which takes a value of -1 in the liquid and +1 in the solid.  In addition, the nondimensionalized temperature is indicated by $U$, 
\begin{equation}
U=\frac{T-T_m}{L/c_p},
\end{equation}
where $T$ is the local temperature, $T_m$ is the melting temperature, $L$ is the latent heat of fusion, and $c_p$ is the specific heat at constant pressure, such that $U=0$ is the nondimensionalized melting temperature (note that, for this particular problem, we do not need to supply values for $T_m$, $L$, and $c_p$). The free energy of the system, $\mathcal{F}$, is expressed as \cite{karma1998quantitative}
\begin{equation}
\mathcal{F}=\int_{V}\left[\frac{1}{2} \left[W(\textbf{n})\right]^2|\nabla \phi|^2+f_{chem}(\phi,U)\right]\,dV,\label{eq:F_DG}
\end{equation}
where $\left[W(\textbf{n})\right]^2$ is the gradient energy coefficient, $\textbf{n}\equiv \nicefrac{\nabla\phi}{|\nabla\phi|}$ is the normal direction to the interface, and $f_{chem}$ is the chemical free energy density.  In this formulation, $f_{chem}$ is a double-well potential with a simple polynomial formulation \cite{karma1998quantitative},
\begin{equation}
f_{chem}=-\frac{1}{2}\phi^2+\frac{1}{4}\phi^4
+\lambda U\phi\left[1-\frac{2}{3}\phi^2+\frac{1}{5}\phi^4\right],
\label{eqn:f_chem}
\end{equation}
where $\lambda$ is a dimensionless coupling constant.  The interface thickness and directional anisotropy are controlled by $W(\textbf{n})$, which takes the form $W(\textbf{n})=W_0a(\textbf{n})$ in two dimensions. We use a simple form for $a(\textbf{n})$ to reflect in-plane symmetry \cite{karma1998quantitative},
\begin{equation}
a(\textbf{n})=1+\epsilon_m\cos \left[m(\theta-\theta_0) \right],
\label{eqn:a_hat_n}
\end{equation}
where $m$ is a non-negative integer and $\theta$ is the in-plane azimuthal angle, $\tan\theta=n_y/n_x$; $\theta_0$ is an offset azimuthal angle that specifies the misorientation of the crystalline lattice relative to the laboratory frame of reference (in this case, the $x$-axis in the simulation coordinate system). We take 
\begin{equation}
\lambda=\frac{D\tau_0}{0.6267W_0^2}
\end{equation}
because this choice gives quantitative agreement in the so-called ``thin interface limit'' with sharp interface models of dendritic growth in the limit of vanishing interface kinetics \cite{karma1998quantitative}. 

The time scale for the evolution of $\phi$ and $U$ are similar, so both must be described with time-dependent equations.  The evolution of $\phi$, a non-conserved quantity, is governed by the Allen-Cahn equation \cite{karma1998quantitative},
\begin{equation}
\tau(\textbf{n})\frac{\partial\phi}{\partial t} = -\frac{\delta \mathcal{F}}{\delta \phi},\label{eqn:dphi_dt}
\end{equation}
where the kinetic coefficient $\tau(\textbf{n})=\tau_0\left[a(\textbf{n})\right]^2$. The functional derivative in Eq.\ \ref{eqn:dphi_dt} is given as \cite{karma1998quantitative}
\begin{eqnarray}
\tau(\textbf{n})\frac{\partial\phi}{\partial t} & = & \left[\phi-\lambda U\left(1-\phi^2\right)\right]\left(1-\phi^2\right)+\nabla\cdot\left( \left[W(\textbf{n})\right]^2\nabla\phi\right)\nonumber\\
&&+\frac{\partial}{\partial x}\left[|\nabla\phi|^2W(\textbf{n})\frac{\partial W(\textbf{n})}{\partial \left(\frac{\partial\phi}{\partial x}\right)}\right]
+\frac{\partial}{\partial y}\left[|\nabla\phi|^2W(\textbf{n})\frac{\partial W(\textbf{n})}{\partial \left(\frac{\partial\phi}{\partial y}\right)}\right] 
\end{eqnarray}
for two dimensions.  We remind the reader that although  $a(\textbf{n})$, $W(\textbf{n})$ and $\tau(\textbf{n})$ have a vectorial argument, they are scalar quantities. Furthermore, the time evolution of $U$ is governed by thermal diffusion and the release of latent heat at the interface during solidification  \cite{karma1998quantitative},
\begin{equation}
\frac{\partial U}{\partial t} = D\nabla^2U+\frac{1}{2}\frac{\partial \phi}{\partial t}
\label{eqn:dUdt}
\end{equation}
where $D$ is a thermal diffusion constant.

\subsubsection{Parameterization and simulation conditions \label{ssub:solidification-params}}

This section presents the specific details for the solidification and dendritic growth benchmark problem, including the model parameterization, initial conditions, boundary conditions, and computational domain size. The model is parameterized with dimensionless units, and the parameterization is given in Table \ref{tab:solidification-params}. While the diffuse interface width depends on orientation, it varies between four and five units, where the width is defined as the distance over which $-0.9 < \phi < 0.9$.  As mentioned in the beginning of Sec.~\ref{sec:Model-formulation}, the benchmark problem is formulated for two dimensions.  To further reduce computational cost, we simulate one-quarter of a growing dendrite, as is commonly done in earlier works (e.g., Ref.~\cite{karma1998quantitative,warren1995prediction}). One-quarter of a solid seed with a radius of eight units (with the position of the interface defined as $\phi=0$) and a diffuse interface width of one unit is placed in the lower-left corner of the computational domain, surrounded by liquid.  Initially, the entire system is uniformly undercooled with $U=-0.3$.  This undercooling is chosen to challenge numerical solvers somewhat because it increases the thermal diffusion length and requires a larger computational domain size relative to more negative undercoolings.  We select a square computational domain of $(960 \textrm{ units})^2$, which is two times longer than the long dimension used in Ref.~\cite{karma1998quantitative} for the same model parameterization. The dendritic growth simulations are run to $t=1500$.  This time is chosen because the dendrite will grow significantly but the diffusion field will not extend to the far edges of the computational domain (at which point the results can no longer be compared to the analytical solution, though we note that steady-state tip velocity may not be achieved within this time interval). The solutions are output at sychronization times of 15, 75, 150, 300, 600, 900, 1200, and 1500 for direct comparison of different implementation results.   No-flux boundary conditions are chosen for $\phi$ and $U$ on all domain boundaries.  The choice of time and space discretization are discussed in Section~\ref{sec:Numerical-methods}.  Simulation metrics include the total free energy of the system, the dendrite tip velocity, the area fraction of the solid phase, and microstructural snapshots.

\begin{table}
\centering
\caption{\label{tab:solidification-params} Parameterization for the solidification and dendritic growth benchmark problem.}
\begin{tabular}{c c c}
\hline
Quantity & Symbol & Value \\\hline
Interface thickness & $W_0$ & 1  \\
Rotational symmetry order & $m$ & 4 \\
Anisotropy strength & $\epsilon_4$ &  0.05 \\
Offset angle & $\theta_0$ & 0 \\
Relaxation time & $\tau_0$ & 1\\
Diffusion coefficient & $D$ & 10\\
Undercooling & $\Delta$ & -0.3\\
\hline
\end{tabular}

\end{table}

\subsection{Linear elasticity and equilibrium shape of an elastically constrained precipitate \label{sub:linear-elasticity}}

Elastic stresses arise in myriad materials systems, such as precipitate-strengthened alloys. A coherent precipitate will be elastically stressed if a lattice parameter is mismatched between the matrix and precipitate phases, while volumetric elastic stresses may arise for an incoherent precipitate due to thermal expansion mismatch with the matrix.  The presence of elastic stress within the precipitates affects properties such as phase stability, precipitate shape and alignment/orientation, and precipitate coarsening behavior.  The second benchmark problem targets the physics of linear elasticity by finding the equilibrium shape of a coherent, misfitting, elastically constrained precipitate for which the elastic moduli of each phase are anisotropic and which are different for each phase.  The model parameterization is chosen to explore simulation results for precipitate sizes that are smaller and larger than the size at which shape bifurcation occurs \cite{thompson1994equilibrium}.  Note that we neglect the dynamics of morphological evolution in this problem, because verification of solver accuracy via the equilibrium shape must be performed before studying the added complexity of dynamics. While many phase field models that incorporate elasticity are quite complex, we simplify the model for the benchmark problem as much as possible to keep the test focused on the targeted physics. For this problem, we follow the approach presented in Ref.~\cite{jokisaari2017superalloy}, which in turn is similar to the diffuse-interface model presented in Ref.~\cite{leo1998diffuse}.  

In this problem, one phenomenological order parameter, $\eta$, is evolved, which has a value of 0 in the matrix and a value of 1 in the precipitate for an unstressed system with planar interfaces. This choice makes interpolation of materials properties between phases straightforward. The free energy of the system, $\mathcal{F}$, includes contributions from interfacial and elastic energy and is expressed as 
\begin{equation}
\mathcal{F}=\int_{V}\left(f_{bulk}\left(\eta\right) + \frac{\kappa}{2}|\nabla \eta|^{2} + f_{el}\left(\eta\right) \right)dV,
\label{eq:F_tot}
\end{equation}
where $f_{bulk}$ is the bulk free energy density, $\kappa$ is the gradient energy coefficient, and $f_{el}$ is the local elastic free energy density. The $f_{bulk}$ term is a symmetric double-well with minima of zero, such that its contribution is only to the interfacial energy. As discussed in Ref.~\cite{jokisaari2017superalloy}, we choose $f_{bulk}$ to have a 10th-order polynomial form,
\begin{equation}
f_{bulk}=w\sum_{j=0}^{10}a_j\eta^j,
\label{eq:f_chem}
\end{equation}
which makes the energy wells of the matrix and precipitate phases deep and narrow.  This prevents the actual value of $\eta$ in each phase from shifting significantly from its equilibrium value due to the presence of a curved interface or elastic strain.  The height of the energy barrier is controlled by $w$. The $f_{bulk}$ coefficients are given in Table \ref{tab:fbulk-params}, and ensure that $f_{bulk}\left(0\right)=f_{bulk}\left(1\right)=f_{bulk}'\left(0\right)=f_{bulk}'\left(1\right)=0$ and that the energy curve remains concave down between the two energy wells.

\begin{table}
\centering
\caption{\label{tab:fbulk-params} Parameterization of the $f_{bulk}$ term in Eq.~\ref{eq:f_chem} (from Ref.~\cite{jokisaari2017superalloy}). The large number of significant digits are necessary to ensure that the first derivative of $f_{bulk}$ is zero at $\eta=0$ and $\eta=1$.}
\begin{tabular}{c c c c}
\hline
$a_j$ & parameter value & $a_j$ & parameter value \\\hline
$a_0 = a_1$ & 0  & $a_6$ & 2444.046270 \\
$a_2$ & 8.072789087 & $a_7$ &  -3120.635139\\
$a_3$ & -81.24549382 & $a_8$ &  2506.663551\\
$a_4$ & 408.0297321 & $a_9$ & -1151.003178\\
$a_5$ & -1244.129167 & $a_{10}$ & 230.2006355\\
\hline
\end{tabular}

\end{table}

The elastic energy density is given as \cite{jokisaari2017superalloy}
\begin{equation}
f_{el}\left(\eta\right)=\frac{1}{2}\sigma_{ij} \epsilon_{ij}^{el},
\label{eq:f_el}
\end{equation}
where $\sigma_{ij}=C_{ijkl}\left(\eta\right) \epsilon_{ij}^{el}$ is the elastic stress, $\epsilon_{ij}^{el}$ is the elastic strain, and $C_{ijkl}\left(\eta\right)$ is the elastic stiffness tensor such that the system is mechanically stable (the Einstein summation convention is used).  To incorporate the dependence of the elastic stiffness on the phase, the stiffness is interpolated smoothly from one phase to the other across the diffuse interface, 
\begin{equation}
C_{ijkl}\left(\eta\right)= C_{ijkl}^{matrix}\left[1-h\left(\eta\right)\right]+C_{ijkl}^{precip} \, h\left(\eta\right),
\label{eq:stiffness}
\end{equation}
where $C_{ijkl}^{matrix}$ and $C_{ijkl}^{precip}$ are the stiffness tensors of the matrix and precipitate phases, respectively, and $h\left(\eta\right)=\eta^3\left(6\eta^2-15\eta+10\right)$ is a smooth interpolation function that ensures that $h\left( 0 \right)=h\left( 1 \right)=h'\left( 0 \right)=h'\left( 1 \right)=0$ \cite{leo1998diffuse}. 

Because the lattice parameters of the two phases are different, the elastic strain differs from the total strain, $\epsilon_{ij}^{total}$, as \cite{eshelby1957determination}
\begin{equation}
\epsilon_{ij}^{el}=\epsilon_{ij}^{total}-\epsilon_{ij}^{0}\left(\eta\right),
\label{eq:elastic_strain}
\end{equation}
where $\epsilon_{ij}^{0}$ is the local stress-free strain. It is calculated as 
\begin{equation}
\epsilon_{ij}^0\left(\eta\right)= \epsilon_{ij}^T \, h\left(\eta\right),
\label{eq:misfit_strain}
\end{equation}
where $\epsilon_{ij}^{T}$ is the crystallographic misfit strain tensor between the matrix and precipitate phases defined with respect to the matrix.  Finally, the total strain is related to the displacements, $u_i$, as \cite{eshelby1957determination}
\begin{equation}
\epsilon_{ij}^{total}=\frac{1}{2}\left[\frac{\partial u_i}{\partial x_j}+\frac{\partial u_j}{\partial x_i}\right].
\label{eq:total_strain}
\end{equation}

In this problem, precipitate shapes must evolve to their equilibrium shape while remaining as small particles embedded in a much larger matrix. To do so, we employ the Cahn-Hilliard equation to perform fictive time evolution \cite{leo1998diffuse,jokisaari2017superalloy}, which conserves the total integral of $\eta$ within the simulation. The evolution of $\eta$ is given as 
\begin{equation}
\frac{\partial\eta}{\partial t}=\nabla\cdot\left[M\nabla\left\{ \frac{\delta \mathcal{F}}{\delta\eta}\right\} \right],
\label{eq:CH}
\end{equation}
where $M$ is the mobility and the chemical potential is 
\begin{equation}
\mu \equiv \frac{\delta \mathcal{F}}{\delta\eta}=\frac{\partial f_{chem}}{\partial\eta}+\frac{\partial f_{elastic}}{\partial\eta}-\kappa\nabla^{2}\eta.\label{eq:dFdn}
\end{equation}
We have flexibility in choosing $M$, as we are only interested in the final state of the system. Furthermore, we assume that the relaxation dynamics for elasticity are much faster than for the diffusion of $\eta$, as is generally the case for phase field models.  As such, we solve the time-independent equation for mechanical equilibrium at each time step, 
\begin{equation}
\nabla\cdot\sigma_{ij} = 0. \label{div_stress}
\end{equation}

\subsubsection{Parameterization and simulation conditions \label{ssub:elasticity-params}}

Similar to Section~\ref{ssub:solidification-params}, we present the specific details of model parameterization, initial conditions, boundary conditions, and computational domains for the benchmark problem on linear elasticity.  As with the solidification and dendritic growth problem, this problem is solved in two dimensions to reduce computational costs, but note that we do not utilize symmetry to further reduce the problem size.  The matrix and precipitate phases have cubic symmetry, such that three independent elastic stiffnesses exist for each phase: $C_{1111}$, $C_{1122}$, and $C_{1212}$ \cite{nye1957physical}, and we take $C_{ijkl}^{precip}=1.1C_{ijkl}^{matrix}$.  In addition, the precipitate misfit strain takes the form $\epsilon^T_{11}=\epsilon^T_{22} > 0$, $\epsilon^T_{12}=0$.  Because this benchmark problem relies on the balance between interfacial and elastic energy, we use dimensional units of attojoules and nanometers.  The diffuse interface width is chosen as 5 nm for  $0.05 < \eta < 0.95$ and the interfacial energy is chosen as 50 aJ/nm$^2$ (equivalent to 50 mJ/m$^2$). The model parameters are given in Table~\ref{tab:elasticity-params}.  Simulation metrics include the final total free energy of the system, the final elastic energy density of the system, the final area fraction of the precipitate phase, and the final dimensions of the precipitate as defined by several factors detailed in Section \ref{sub:discussion-elasticity}.

\begin{table}
\centering
\caption{\label{tab:elasticity-params} Parameterization for the elastically  constrained precipitate benchmark problem.  Note that 1 aJ/nm$^3$ is equivalent to 1 GPa.}
\begin{tabular}{c c c}
\hline
Quantity & Symbol & Value \\\hline
Gradient energy coefficient & $\kappa$ & 0.29 aJ/nm \\
Well height & $w$ & 0.1 aJ/nm$^3$ \\
Mobility & $M$ & 5 \\
Misfit strain & $\epsilon^T_{11}=\epsilon^T_{22}$ &  \SI{0.5}{\percent} \\
Elastic stiffness of matrix & $C^{matrix}_{1111}$, $C^{matrix}_{1122}$, $C^{matrix}_{1212}$ & (250, 150, 100) aJ/nm$^3$ \\
Elastic stiffness of precipitate & $C^{precip}_{1111}$, $C^{precip}_{1122}$, $C^{precip}_{1212}$ & (275, 165, 110) aJ/nm$^3$ \\
\hline
\end{tabular}

\end{table}

We utilize both circular and elliptical initial precipitate shapes for a given initial precipitate area \cite{thompson1994equilibrium,li2004two}; all initial precipitates have a diffuse interface width of 5 nm. To have an equal area for an ellipse as a circle with radius $r$, we choose ellipse axes as $a_{[10]}=r/0.9$ and $a_{[01]}=0.9r$. Simulations are performed for two initial precipitate sizes: a smaller one with an area of $20^2 \pi \textrm{ nm}^2$ and a larger one with an area of $75^2 \pi \textrm{ nm}^2$. The center of each precipitate is embedded in the center of a square computational domain, which is given the coordinate (0,0).  The computational domain is $(400 \textrm{ nm})^2$ for the smaller precipitates and $(1500 \textrm{ nm})^2$ for the larger precipitates to allow long-range elastic fields to decay.  No-stress boundary conditions are applied for the displacements, and no-flux boundary conditions are applied for $\eta$. Because our implementation is based on solving for displacements rather than strain, we specify $u_{[10]}=0$ at the top, middle, and bottom of the $y=0$ axis ($e.g.,$ in the [01] direction) and $u_{[01]}=0$ at the top, middle, and bottom of the $x=0$ axis ($e.g.,$ in the [10] direction) to remove the nullspace in the solution.  Simulations are run until equilibrium is achieved.

As noted in Sec.~\ref{sub:linear-elasticity} (and discussed in Ref.~\cite{jokisaari2017superalloy}), the presence of elastic strain energy or a curved interface will increase the final value of $\eta$ in both the matrix and precipitate phases from the equilibrium value given by the common tangent of $f_{bulk}$.  In addition, the precipitate may change size during the course of the energy relaxation because of the shifting balance between the $f_{bulk}$ and $f_{el}$ energy contributions.  Because the precipitate volume within the computational domain is much smaller than that of the matrix, a precipitate may shrink entirely away in the process achieving the equilibrium value of $\eta$ in the matrix.  To avoid this, the initial value of $\eta$ in the matrix should be set slightly greater than zero.  For the simulations with the small particles, we set $\eta^{matrix}=0.0065$, while for the large particles, $\eta^{matrix}=0.005$. In addition, we set $\eta^{precip}=1$ for all simulations.

\section{Numerical methods \label{sec:Numerical-methods}}

In addition to presenting the problems themselves and specifying relevant output to be compared between different approaches to solving the problems, we also provide our own example solutions. Our solutions are not part of the benchmark problems themselves, but serve as walk-through examples discussing some issues that may arise and that may be generic to any numerical implementation.
%
For this set of benchmark problems we use a bespoke, in-house code and an application based on the MOOSE computational framework \cite{gaston2014continuous,gaston2015physics}.  MOOSE-based simulations are performed for both problems, while the bespoke code is used only for the first problem on solidification and dendritic growth.  We have also used MOOSE to provide example solutions for the first set of benchmark problems \cite{jokisaari2017benchmark}.  

The numerical methods in the bespoke code are extremely similar to those used in many of the phase field studies within the literature on dendritic growth, and serve as a standard against which to compare the results of the MOOSE-based simulations.  The bespoke code employs finite differencing on a fixed mesh for spatial discretization and operator-split explicit Euler time stepping with a fixed time step size; it is parallelized using OpenMP. The grid spacing is equal in the [10] and [01] directions and a skewed nine-point stencil for the Laplacian operator is used.  Following Ref.~\cite{karma1998quantitative}, we select a discretization size, $dx$, of 0.4 units.  A convergence study is performed to determine the time step size for the bespoke code by comparing the free energy evolution through $t \approx 10$.  The maximum stable time step size, $dt$, is $\approx$ 0.009. The energy evolution converges as $dt$ approaches 0.001; we choose $dt=0.003$ for a balance between integration error and simulation run time. 

MOOSE is an open-source, fully coupled, fully implicit finite element solver framework with adaptive time stepping and adaptive meshing capabilities.  For the simulations in this work, the finite element meshes are comprised of square, four-node quadrilateral elements and linear Lagrange shape functions are employed for all nonlinear variables. Meshes are generated with the internal MOOSE mesh generator.  As in the first benchmark paper, the Cahn-Hilliard equation (Eq.\ \ref{eq:CH}) in the linear elasticity problem is split into two second-order equations \cite{elliott1989second,tonks2012object} to avoid computationally expensive fourth-order derivative operators. The system of nonlinear equations is solved with the preconditioned Jacobian Free Newton-Krylov (PJFNK) \cite{knoll2004jacobian} method.  For the dendritic growth problem, we employ the additive Schwarz method for preconditioning and LU factorization for sub-preconditioning, while for the elastically constrained precipitate problem, we use KSP preconditioning and LU factorization for sub-preconditioning.  The dendritic growth simulations are solved with a nonlinear relative tolerance of $1\times10^{-8}$ and a nonlinear absolute tolerance of $1\times10^{-11}$, while the elastically constrained precipitate simulations are solved with with a nonlinear relative tolerance of $1\times10^{-6}$ and a nonlinear absolute tolerance of $1\times10^{-9}$.

Adaptive meshing and adaptive time stepping are employed to reduce the computational cost of the finite element simulations. The dendritic growth simulations employ minimum and maximum element sizes of 0.4 units and 12.8 units, respectively, corresponding to five levels of mesh refinement/coarsening.  The elastically constrained precipitate simulations have a minimum mesh size of 1.0 nm, with a maximum of four and five levels of coarsening for the smaller and larger precipitate simulations, respectively.  Gradient jump indicators \cite{kirk2006libmesh} are employed for all nonlinear variables to determine mesh adaptivity.  The elements with the lowest 2\% of error for $\phi$ and $U$ and 5\% of error for $\eta$, $u_{[10]}$, and $u_{[01]}$ are chosen for coarsening, while the elements with the largest 75\% error for $\phi$, $U$, and $\eta$ and the largest 50\% of error for $u_{[10]}$ and $u_{[01]}$ are chosen for refinement.  

Several different time integrators are employed in the MOOSE-based simulations. We choose to investigate four different time integration schemes for the dendritic growth simulations, namely the first-order implicit Euler (IE) scheme and the second-order Crank-Nicolson (CN), second backward differentiation formula (BDF2) \cite{iserles2009first}, and L-stable, two-stage diagonally implicit Runge-Kutta (DIRK) \cite{alexander1977diagonally} schemes; only the BDF2 scheme is applied for the elastically constrained precipitate simulations.  Implicit timestepping allows larger time steps to be taken versus explicit time stepping, but due to the iterative nature of implicit solvers, each time step is more computationally intensive.  In addition, the MOOSE-based simulations employ the ``IterationAdaptive'' time stepper, which attempts to maintain a constant number of nonlinear iterations.  This time stepper can greatly reduce computational time because it increases the time step as the driving force to evolve the system decreases. For the dendritic growth simulations, we target a time step size $dt=0.3$, two orders of magnitude larger than that used in the explicit time integration. We choose the adaptivity parameters that allow the time step to grow quickly from the initial $dt=0.003$ and then remain fixed at $dt=0.3$ except when a synchronization time is reached.  The timestep is allowed to grow by 5\% to reach the target $dt$ and the target number of nonlinear iterations per time step is four. The same time stepper and $dt$ growth factor are used for the elastically constrained precipitate simulations, but no maximum value is set for $dt$; a target of six nonlinear iterations with a window of $\pm 1$ iterations is chosen.  

\section{Results and discussion \label{sec:results-and-discussion}}

In this section, we demonstrate the sensitivity and utility of the benchmark problems by evaluating simulation results when minor, ostensibly negligible, algorithm and model details are varied.  We study different implicit time integration algorithms for the solidification and dendritic growth problem and explore the effect of the initial precipitate shape and elastic moduli approximations on elastic bifurcation behavior for the elastically constrained precipitate problem.  By providing a basis for comparison, these problems will help the growing phase field community in developing quantitative materials models.  The input files, data, and code are available in the Materials Data Facility (\url{https://doi.org/10.18126/m2qs6z}).

\subsection{Solidification and dendritic growth
\label{sub:discussion-dendritic-growth}}

For this benchmark problem, we study the effect that different implicit time integrators have on the simulation results. Explicit time integration methods enforce a maximum $dt$ as a function of spatial discretization, $dx$, due to numerical instability; generally it is assumed that integration error is small because of this limitation \cite{cheng2007controlling}.  However, many implicit time integrators are unconditionally stable, removing the naive link between $dt$ and $dx$; larger time step sizes may be taken at the risk of unacceptably large integration error.  Rather than taking a mathematical approach to study integration error, we perform ``numerical experiments'' to understand how simulation results such as dendrite tip velocity, dendrite shape, and free energy evolution are impacted.  
%
%
A standardized problem, such as the benchmark problem presented in this work, is a good way to evaluate a new algorithm or solver for a problem that is very sensitive to instabilities, such as dendritic growth,  before delving into more complex analysis.

\begin{figure}\centering




\subfloat[\label{sfig:dendrite_EE}]{\includegraphics[scale=0.31]{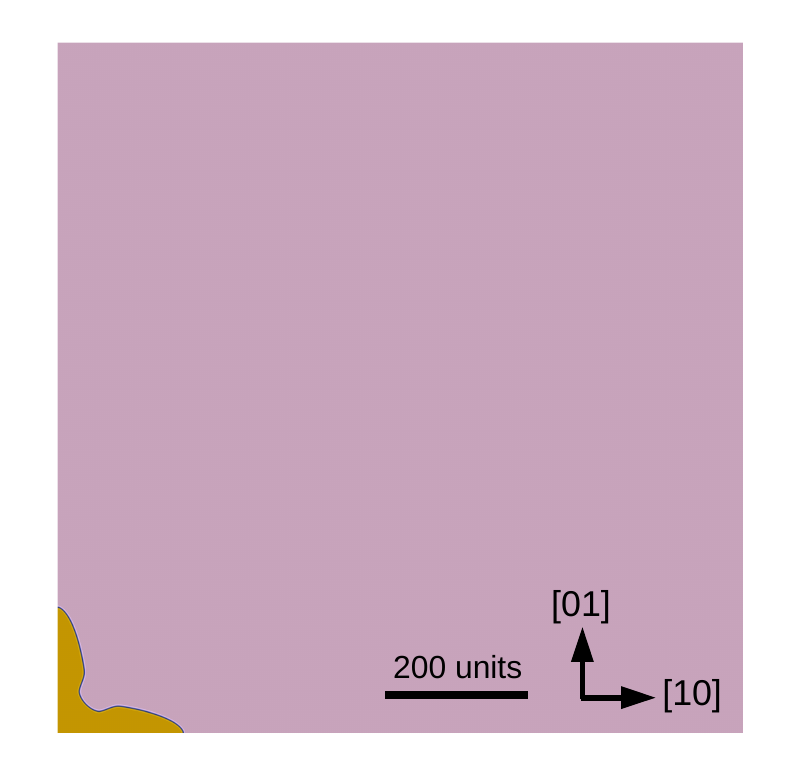}} 
\subfloat[\label{sfig:dendrite_overlay}]{\includegraphics[scale=0.3]{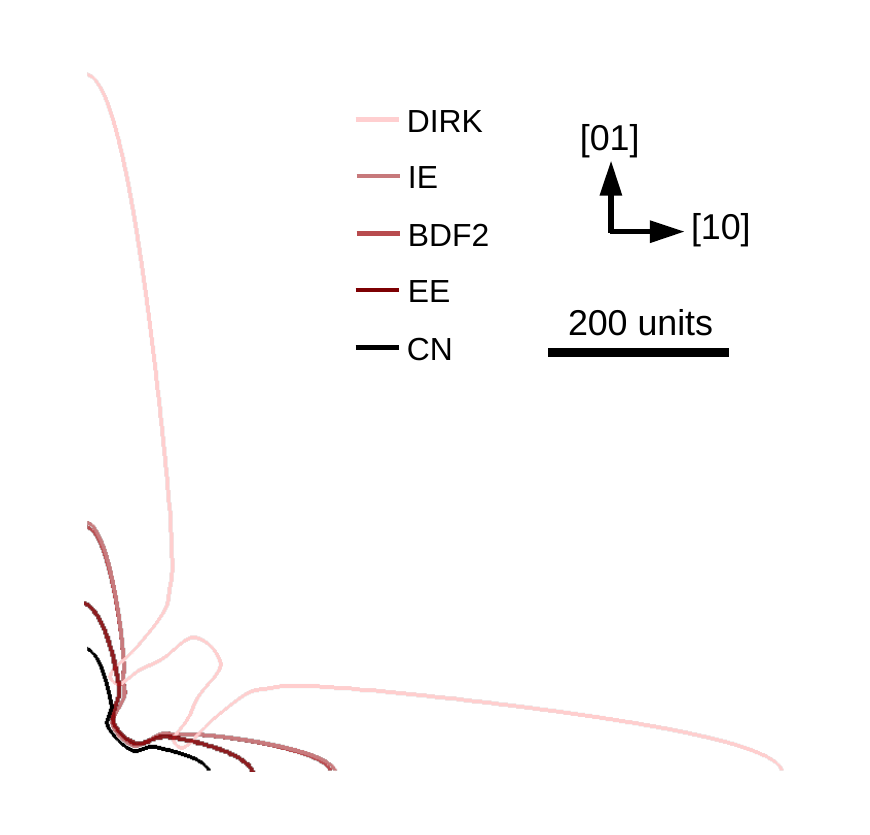}}

\caption{The morphology of dendrites simulated to $t=1500$ (except for DIRK, which is to $t=1146$).  (a) An example of the full computational domain and dendrite, shown for the bespoke code (finite difference spatial discretization and explicit Euler time integration).  (b) A comparison of the final dendrite morphologies simulated by DIRK, implicit Euler, BDF2, explicit Euler, and Crank-Nicolson time integrators.  Contours mark the $\phi=0$ level set computed by each method.\label{fig:dendrites_comparison}}
\end{figure}

\begin{figure}\centering
\subfloat[\label{sfig:dendrite_volfrac}]{\includegraphics[scale=1]{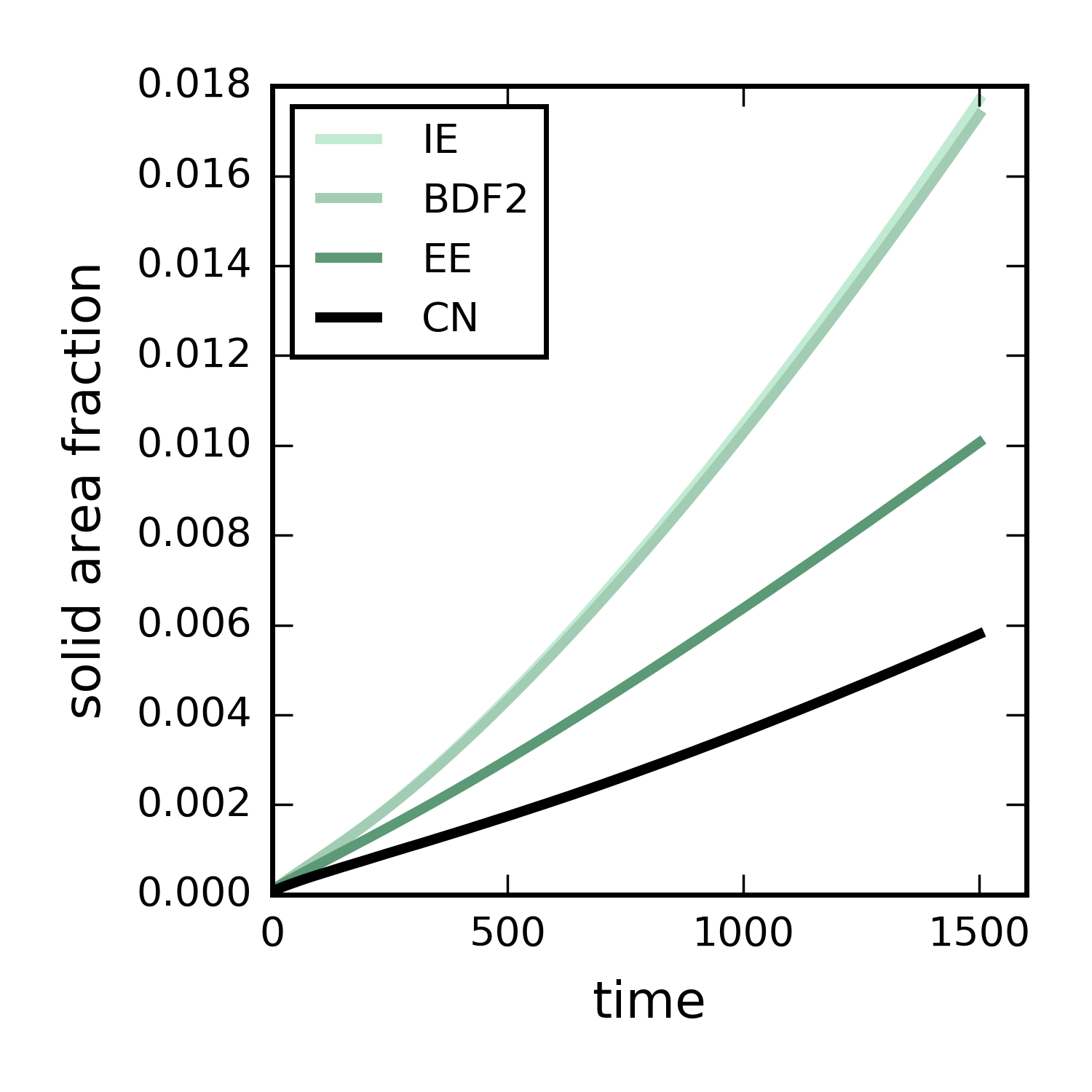}} \hfill \subfloat[\label{sfig:dendrite_energy}]{\includegraphics[scale=1]{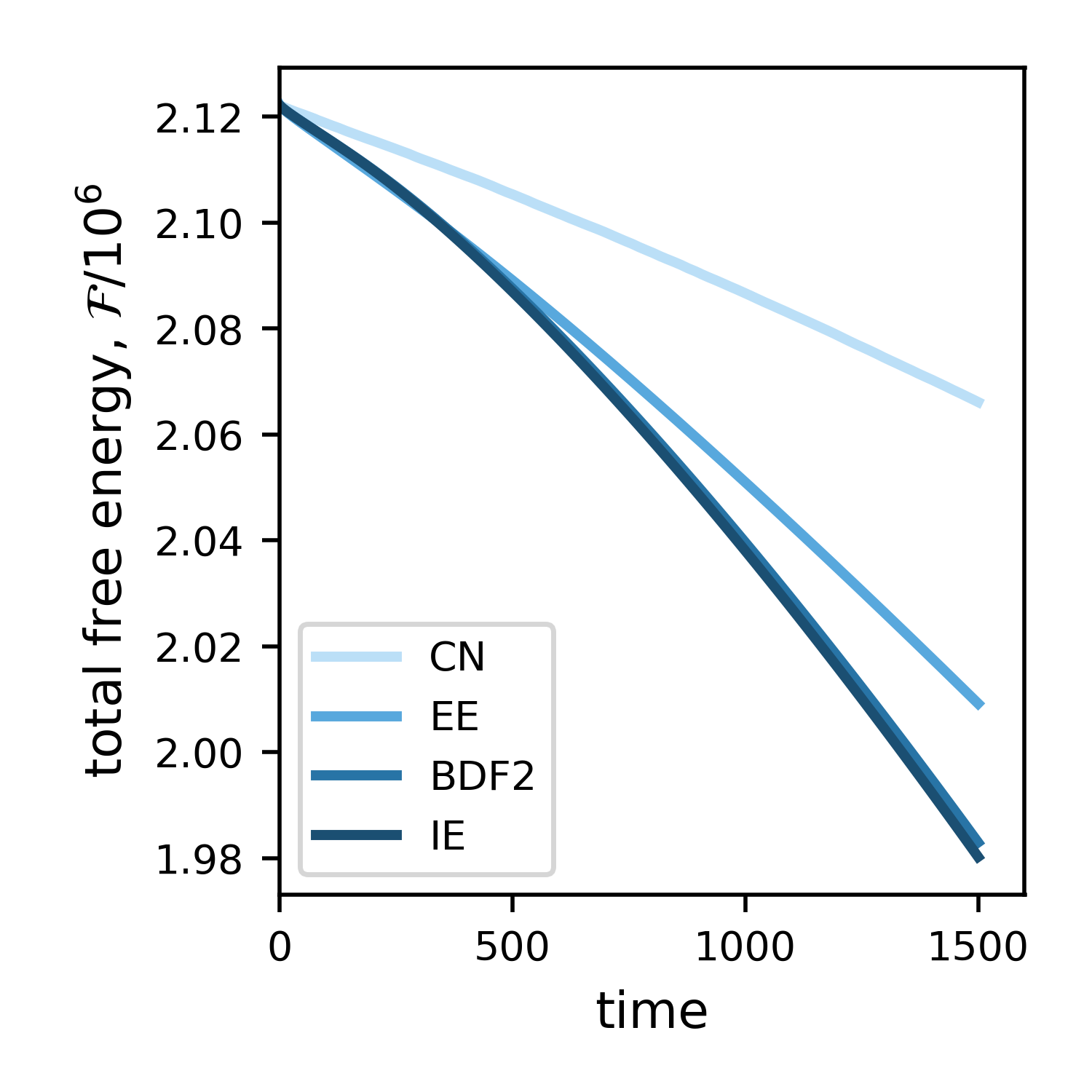}}

\caption{The (a) area of the solid phase and (b) total system energy for dendritic growth simulated with the explicit Euler, implicit Euler, BDF2, and Crank-Nicolson  time integrators. Note that the curves for BDF2 and implicit Euler almost perfectly overlap.  \label{fig:dendrites_energy_area}}
\end{figure}

We choose the total free energy of the system, the dendrite tip velocity, the area fraction of the solid phase, and microstructural snapshots as simulation metrics; these metrics are chosen because they are relatively easy to calculate and compare and their variety should provide sensitivity to probe a variety of different algorithms and implementations.  Figure~\ref{fig:dendrites_comparison} shows the final microstructure at $t=1500$ for the dendrites simulated with the bespoke code as well as the MOOSE-based simulations with implicit Euler (IE), BDF2,   Crank-Nicolson (CN), and DIRK (to $t=1146$)  schemes.  Note that the DIRK simulation was terminated after $t=1146$ for exceeding the alloted wall time. There are clear differences in the size of the dendrites, and in the case of the DIRK time integrator, the evolution is so extensive that an additional branch forms at $45^{\circ}$ between the two primary dendrite arms. In addition, the evolution of the dendrite volume fraction and the evolution of the total system energy are shown in Fig.~\ref{fig:dendrites_energy_area}.  The evolution of the dendrite area fraction shows clear differences depending on which time integrator is employed.  The area fraction for the simulations with BDF2 and IE are almost identical, while the area fraction simulated via BDF2 is 72\% larger than that simulated with the bespoke code at $t=1500$. Conversely, the area fraction simulated via CN is 42\% smaller than that simulated with the bespoke code.  However, the DIRK result is vastly different, with the dendrite increasing in area much faster than that simulated with any other method.  Discussion with the MOOSE developers indicates that the DIRK implementation may not correctly handle the coupled time derivative in Eq.~\ref{eqn:dUdt}.  As such, we will not further consider DIRK results. 

The dendrite tip velocity is a different measure of the system evolution. As mentioned in Sec.~\ref{sub:solidification}, solutions calculated via solvability theory (also termed the Green's function method or the boundary integral method in the literature) in Ref.~\cite{karma1998quantitative} for comparison. Note that the solvability theory result is a steady-state solution, such that a) comparison with phase field results is clearly invalid when the  thermal diffusion field impinges on the domain boundary and b) the dendrite tip velocity goes through a transient before reaching steady-state.  In this work, the tip position, designated by $\phi=0$, is determined along the [10] direction (the bottom edge of the computational domain).  For the finite difference result, the position is found by linear interpolation between three grid points that bracket the $\phi=0$ value.  The velocity is calculated by backward differencing with $\Delta t \approx 10$ (the exact $\Delta t$ is dependent upon whether time adaptivity is used or not).

\begin{figure}\centering

\subfloat[\label{sfig:tip_veloc}]{\includegraphics[scale=1]{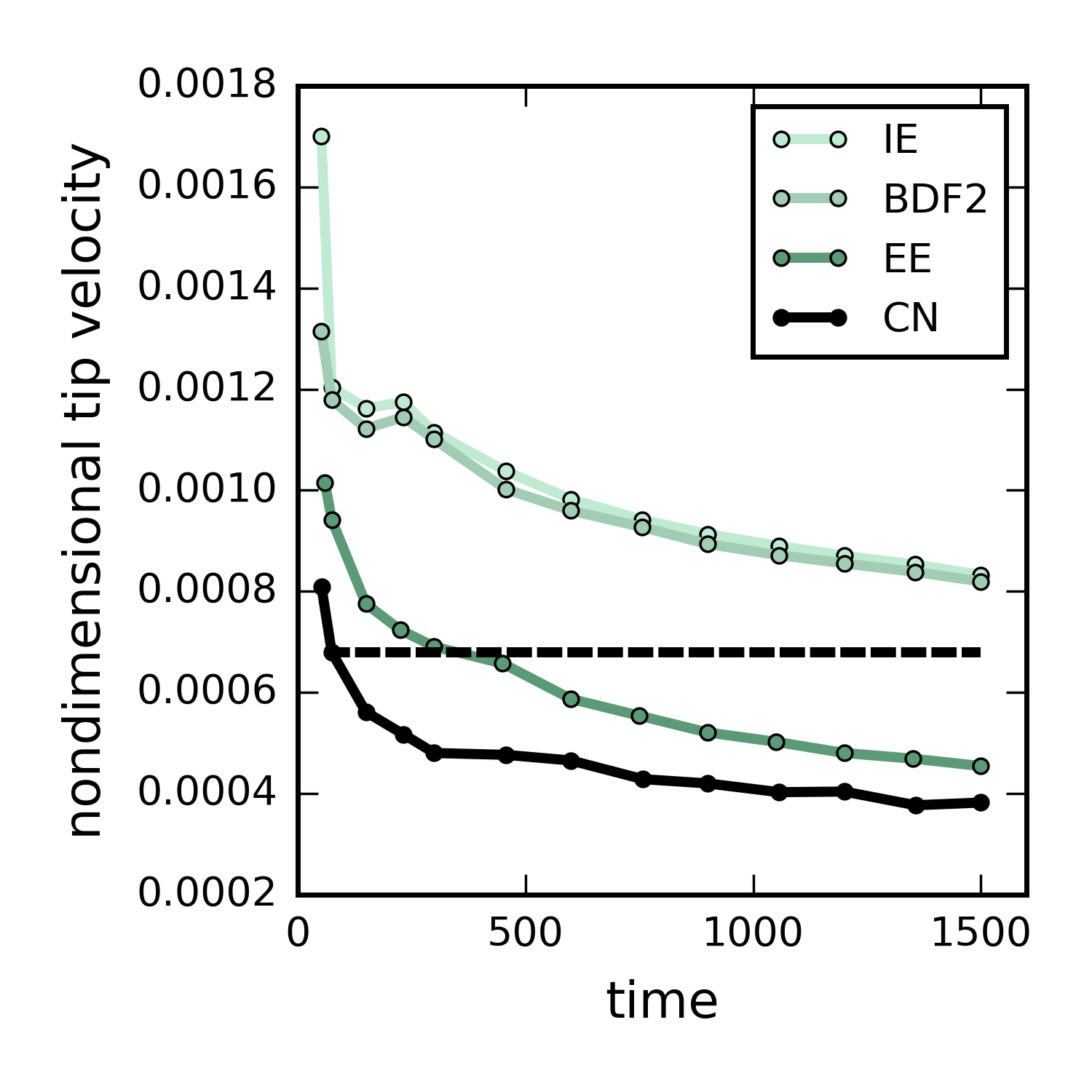}} 
\subfloat[\label{sfig:dendrite_heat}]{\includegraphics[scale=1]{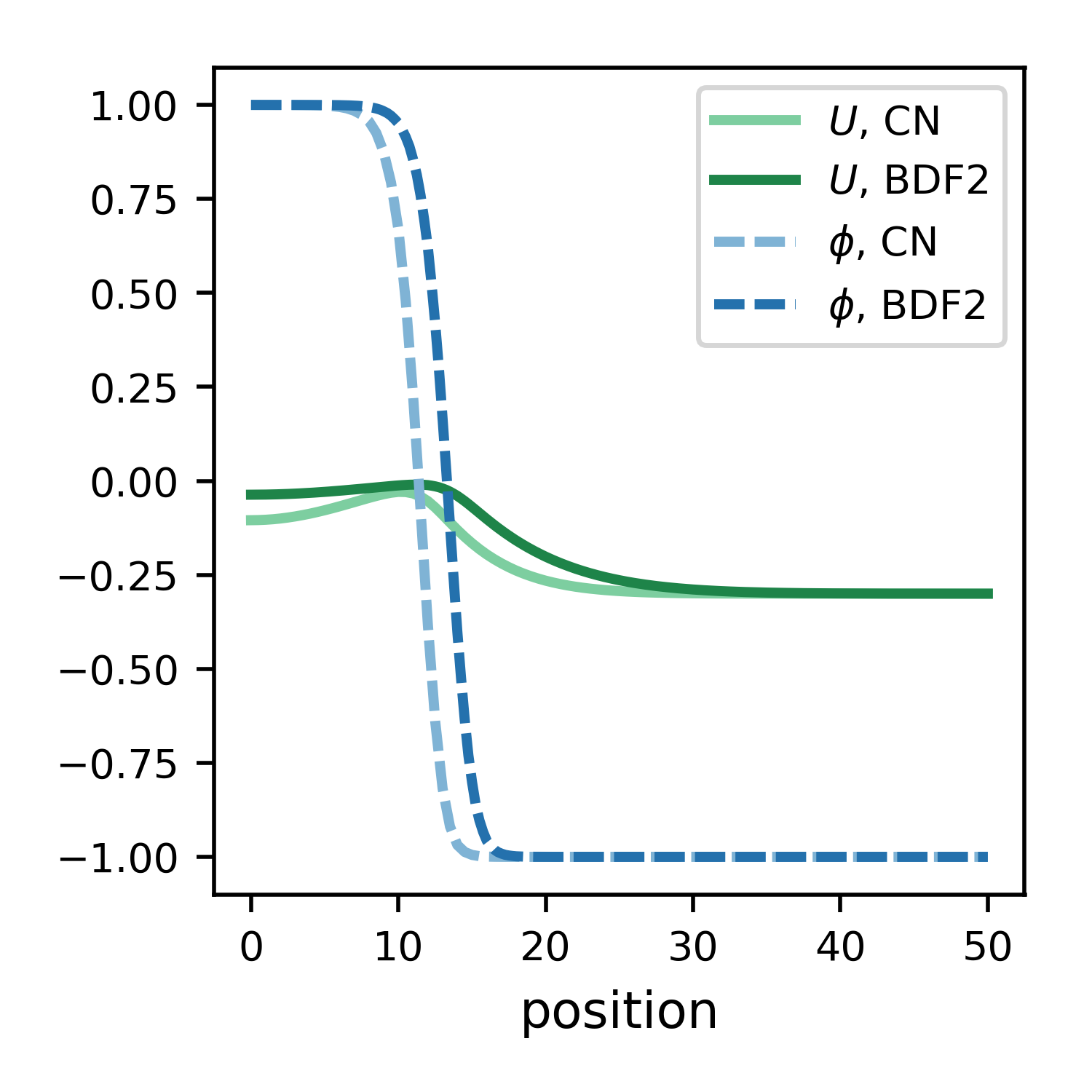}} 

\caption{(a) The non-dimensionalized \cite{karma1998quantitative} tip velocity for dendrites simulated with the explicit Euler, implicit Euler, BDF2, and Crank-Nicolson  time integrators. The straight dashed line indicates steady state sharp-interface (also called the Green's function or boundary integral formulation) solution.  (b) Line cuts of $\phi$ and $U$ along the length of the bottom dendrite arm at $t=6$ for BDF2 and Crank-Nicolson time integrators.} \label{fig:tip_velocity}
\end{figure}

As seen in Fig.~\ref{fig:tip_velocity}, the tip velocity of the dendrites varies depending on the choice of time integrator.  The tip velocities in all simulations are  rapid in the beginning of the simulation and converge toward steady-state and are on the same order of magnitude as that calculated via solvability theory.  However, the velocity simulated via the bespoke code is slower than that calculated via solvability theory.  The velocity simulated via CN is even slower, but the two velocities appear to converge toward the same value.  Conversely, the velocities simulated via IE and BDF2 are almost identical and larger than the value calculated by solvability theory; they appear to be converging to that value.  

\begin{figure}\centering

\subfloat[\label{sfig:bespoke_tip_vel}]{\includegraphics[scale=1]{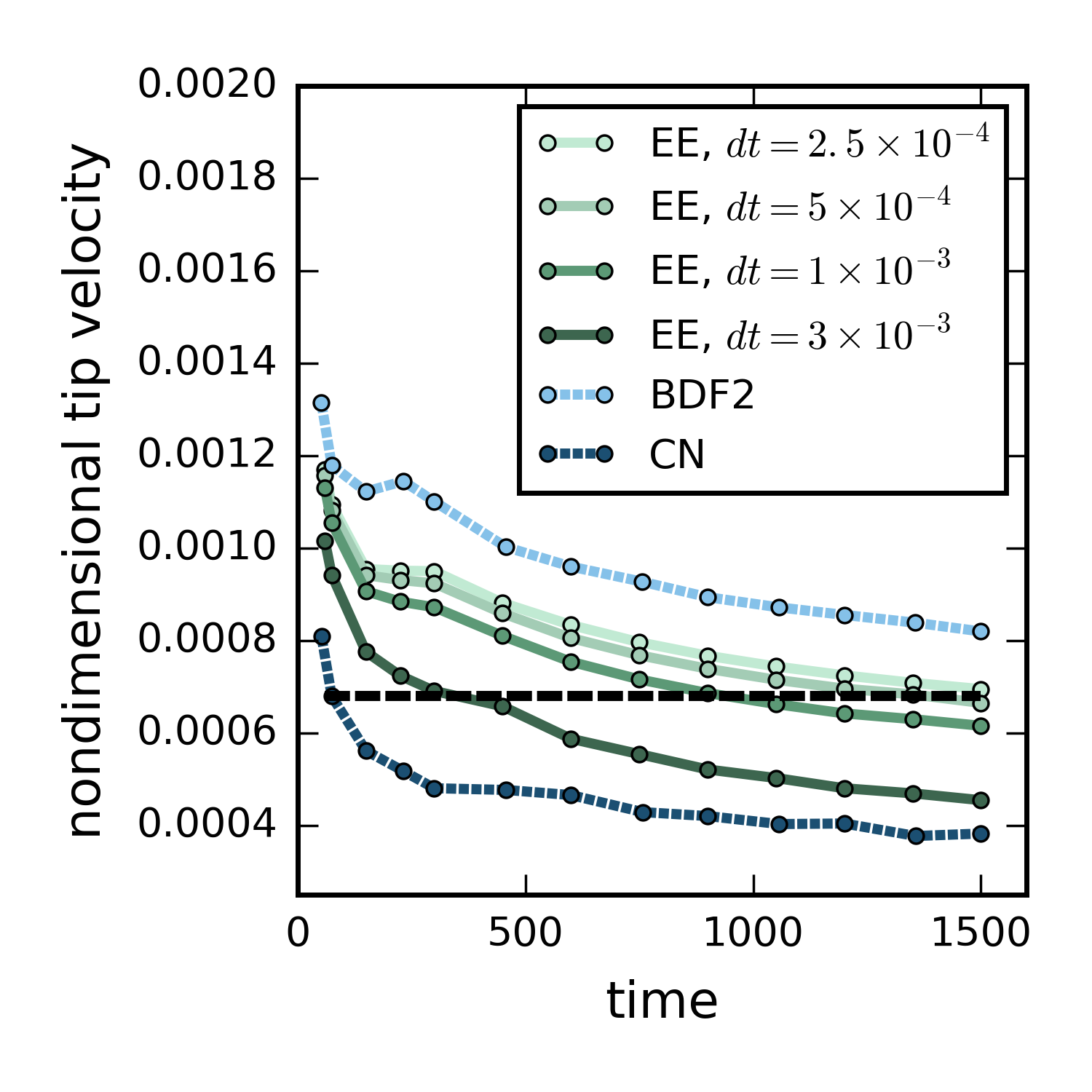}} 
\subfloat[\label{sfig:tiny_dt_freeEnergy}]{\includegraphics[scale=1]{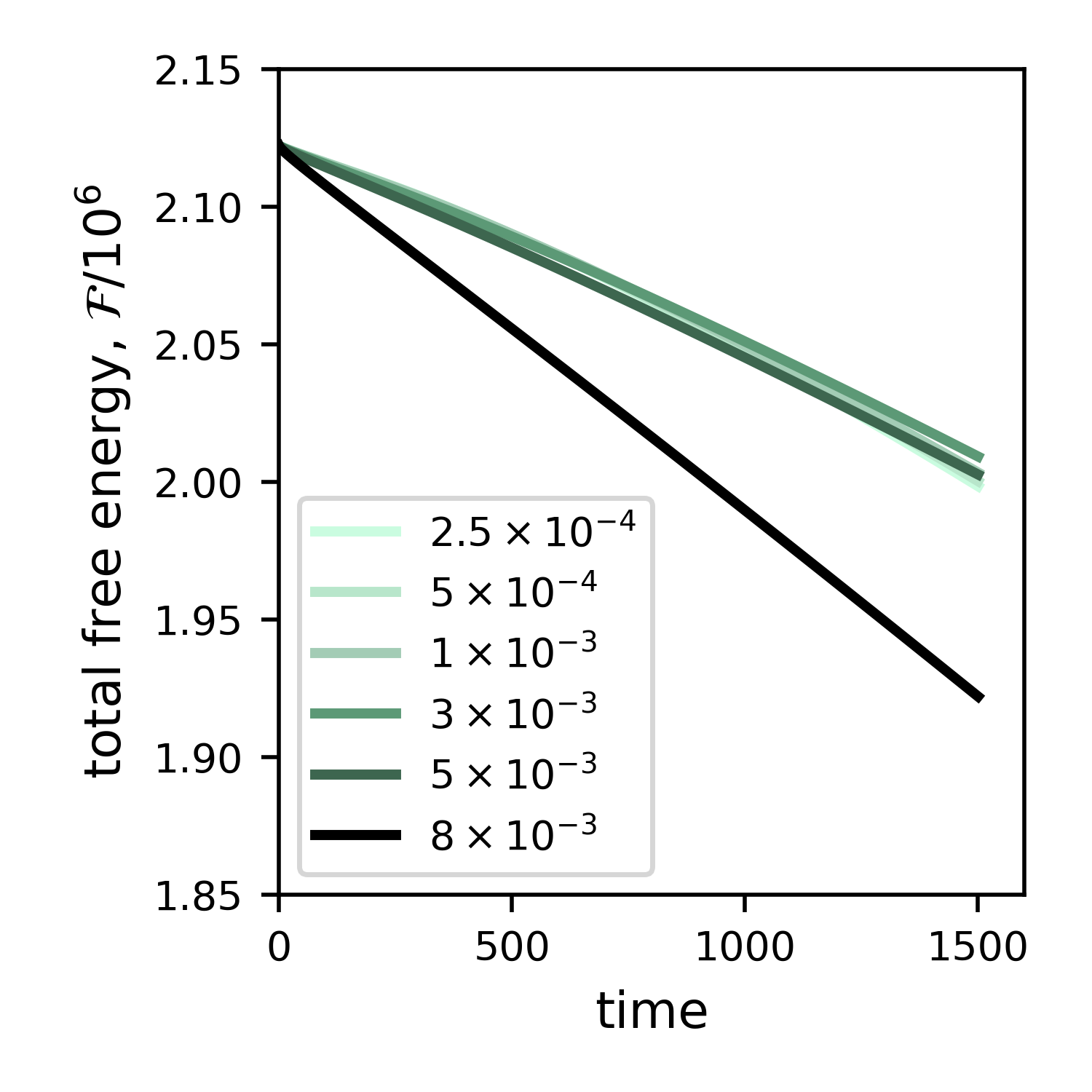}} 

\caption{(a) The non-dimensionalized \cite{karma1998quantitative} tip velocity for dendrites simulated with the bespoke code using different $dt$, overplotted with the BDF2 and CN data in Fig.~\ref{sfig:tip_veloc}. The straight black line indicates the steady state sharp-interface solution. (b) The evolution of the free energy for the system simulated with the bespoke code for different $dt$.} \label{fig:additional_convergence}

\end{figure}

Given the significant scatter in the microstructure and tip velocity simulated by the different methods, and especially the disagreement between the results of the bespoke code and the sharp-interface solution, we revisit our choice of $dt$ and the use of the free energy as a convergence criterion.  We perform several simulations to $t=1500$ with the bespoke code using successively smaller $dt$ and calculate the tip velocity.  The comparison of tip velocity and free energy are shown in Fig.~\ref{fig:additional_convergence}, indicating that the tip velocity converges toward the sharp-interface value as $dt \rightarrow 0.0001$.  Notably, the free energy curves obtained with $dt \leq 0.005$ are all very similar to each other, with the greatest difference being their curvature as a function of time.  As such, the free energy is not an appropriate metric for testing the convergence behavior of this problem and can in fact be completely misleading; relatively small differences in free energy indicate large variations in microstructure evolution.  Nor do we recommend the evolution of the solid area fraction as a convergence criterion, as the simulation must run for considerable time for differences to become evident.  As such, we recommend only the tip velocity as a convergence metric because it is much more sensitive to variations in the result and differences in kinetics are immediately apparent.

We consider the results simulated with the bespoke code using $dt=0.00025$ to be the most accurate solution to the benchmark problem because of the simple numerical algorithms with known error behavior and the observed convergence of the tip velocity toward the sharp-interface solution.  Additional testing with $dt=0.003$ for the MOOSE-based implicit time integration simulations (i.e., with $dt$ two orders of magnitude smaller) indicate that the BDF2 results are converged with $dt=0.3$, while the tip velocity decreases for both the CN and IE results.  However, the tip velocity decreases by only 2\% at $t=75$ using IE and approaches that simulated with BDF2, but decreases by 20\% with the CN algorithm.  These results indicate that the IE and BDF2 results converge, seemingly toward a slightly higher value than that obtained via the bespoke code. 

The variation in tip velocity is explained by the evolution of $U$.  At the start of the simulation, the dendrite is in a uniformly undercooled thermal field; as the dendrite grows, heat is evolved from the moving interface, causing the dendrite to approach the melting temperature and warming the surrounding liquid. Figure~\ref{sfig:dendrite_heat} shows line cuts of $U$ and $\phi$ taken through the length of the horizontal dendrite arm (i.e., the bottom boundary of the computational domain) at $t=6$ for simulations with CN and BDF2 time integration and $dt=0.3$.  The dendrite heats faster with BDF2 and slower with CN; the diffusion of heat into the liquid also appears to be greater with BDF2.  Because the dendrite growth rate is controlled by undercooling and the diffusion of heat, microstructure evolution kinetics are impacted.  

Thus, the broad variation in the simulated kinetics result either from an intrinsic property of the time integration algorithms themselves, or some implementation issue. We point out that the variation in results is not a failure of the benchmark problem, but a feature: the formulation is such that the simulated results are sensitive to variations within the solver algorithms. While beyond the scope of this work, further studies in error analysis and numerical convergence with $dt$ can shed additional light on the results.

\subsection{Linear elasticity and constrained precipitate
\label{sub:discussion-elasticity}}

The elastically constrained precipitate benchmark problem is designed to probe the physical phenomenon of bifurcation, that is, elastic shape instability \cite{thompson1994equilibrium}. It has been shown that the lowest-energy shape of an elastically stressed precipitate with isotropic interfacial energy and cubic crystal symmetry for the matrix and precipitate may not be cuboidal. Instead, the equilibrium shape varies with the size of the precipitate, with the transition from cuboidal symmetry at small size to lower symmetry at larger size termed the shape bifurcation \cite{thompson1994equilibrium}.  A useful tool for analyzing the bifurcation behavior of a precipitate is the $L'$ parameter \cite{li2004two},
\begin{equation}
L' \equiv \frac{\overline{g_{el}} \, l}{\Gamma},
\label{eq:Lprime}
\end{equation}
which is a non-dimensional length that characterizes the ratio of the precipitate's elastic energy to its interfacial energy.  In Eq.~\ref{eq:Lprime}, $\overline{g_{el}}$ is a characteristic dimensional strain energy density, $l$ is a characteristic length of the precipitate equal to the radius of a sphere of the same volume, and $\Gamma$ is the interfacial energy per unit area (the same relationships may also be applied in two dimensions). Thus, precipitates of different absolute sizes can be compared on the same scale, and differences in precipitate morphology at the same $L'$ value can be understood in terms of intrinsic materials properties, e.g., crystal symmetries, elastic stiffness variations, and anisotropies within the interfacial energy and misfit strain.

This benchmark problem studies how the simulated equilibrium morphology of the precipitate is affected by physical factors (precipitate size), model approximations (heterogeneous moduli for the phases, or the homogeneous modulus approximation), and simulation choices (initial precipitate shape). We select the total free energy of the system, the area fraction of the precipitate phase, and factors describing the final precipitate morphology as simulation metrics.  Eight simulations are performed for this benchmark: two different precipitate sizes, with homogeneous moduli or heterogeneous moduli for matrix and precipitate for each size, and with an initial precipitate in the shape of a circle or an ellipse for each moduli case.  Following Ref.~\cite{li2004two}, we calculate $\overline{g_{el}}=2.61\times 10^6 \textrm{ J/m}^3$ for heterogeneous moduli and $\overline{g_{el}}=2.5\times 10^6 \textrm{ J/m}^3$ for homogeneous moduli, a difference of approximately 4\%.  Tables~\ref{tab:elasticity_info_small} and \ref{tab:elasticity_info_big} lists the values of $\int_V f_{el} dV$, $\mathcal{F}$, $L'$ and the precipitate lengths in the [10], [01], and diagonal directions (denoted as $a_{10}$, $a_{01}$, and $a_d$ respectively), where the diagonal angle $\theta_d$ is defined as $\tan \theta_d = a_{01}/a_{10}$.  All dimensions are measured using $\eta=0.5$ to denote the position of the interface.
 
\begin{table}
\centering
\caption{\label{tab:elasticity_info_small} Equilibrium results of the elastically constrained precipitate benchmark problem for the precipitate with an initial area of $20^2 \pi$. The units of length and energy are nm and aJ, respectively.} 
\begin{tabular}{c | c || c c c c c c}
\hline
\multirow{2}{*}{Moduli} & \multirow{2}{*}{Initial shape} & \multicolumn{6}{c}{ Initial area: $20^2\pi$ } \\
& & $L'$ & $a_{10}$ & $a_{01}$ & $a_d$ & $\int_V f_{el} dV$ & $\mathcal{F}$ \\
\hline\hline
\multirow{2}{*}{$C_{ijkl}^{matrix}\neq C_{ijkl}^{precip}$} & circle & 1.02 & 19.9 & 19.9 & 20.6 & 8.350 & 19.39 \\
& ellipse & 1.02 & 19.9 & 19.9 & 20.6 & 8.351 & 19.39 \\
\hline
\multirow{2}{*}{$C_{ijkl}^{matrix} = C_{ijkl}^{precip}$} & circle & 0.99 & 20.1 & 20.1 & 20.9  & 8.225 & 19.03 \\
& ellipse & 0.99 & 20.1 & 20.1 & 20.9 & 8.225 & 19.03 \\
\hline
\end{tabular}
\end{table}

\begin{table}
\centering
\caption{\label{tab:elasticity_info_big} Equilibrium results of the elastically constrained precipitate benchmark problem for the precipitate with an initial area of $75^2 \pi$. The units of length and energy are nm and aJ, respectively.} 
\begin{tabular}{c | c || c c c c c c}
\hline
\multirow{2}{*}{Moduli} & \multirow{2}{*}{Initial shape} & \multicolumn{6}{c}{ Initial area: $75^2\pi$ } \\
& & $L'$ & $a_{10}$ & $a_{01}$ & $a_d$ & $\int_V f_{el} dV$ & $\mathcal{F}$ \\
\hline\hline
\multirow{2}{*}{$C_{ijkl}^{matrix}\neq C_{ijkl}^{precip}$} & circle & 3.90 & 70.7 & 70.7 & 57.6 & 118.8 & 184.4 \\
& ellipse & 3.90 & 77.9 & 63.8 & 82.3 & 118.6 & 184.4 \\
\hline
\multirow{2}{*}{$C_{ijkl}^{matrix} = C_{ijkl}^{precip}$} & circle & 3.78 & 71.0 & 71.0 & 83.2 & 116.5 & 179.3 \\
& ellipse & 3.79 & 100.9 & 50.6 & 93.9 & 115.1 & 179.0 \\
\hline
\end{tabular}
\end{table}

\begin{figure}\centering

\subfloat[\label{sfig:small_IC}]{\includegraphics[scale=0.3]{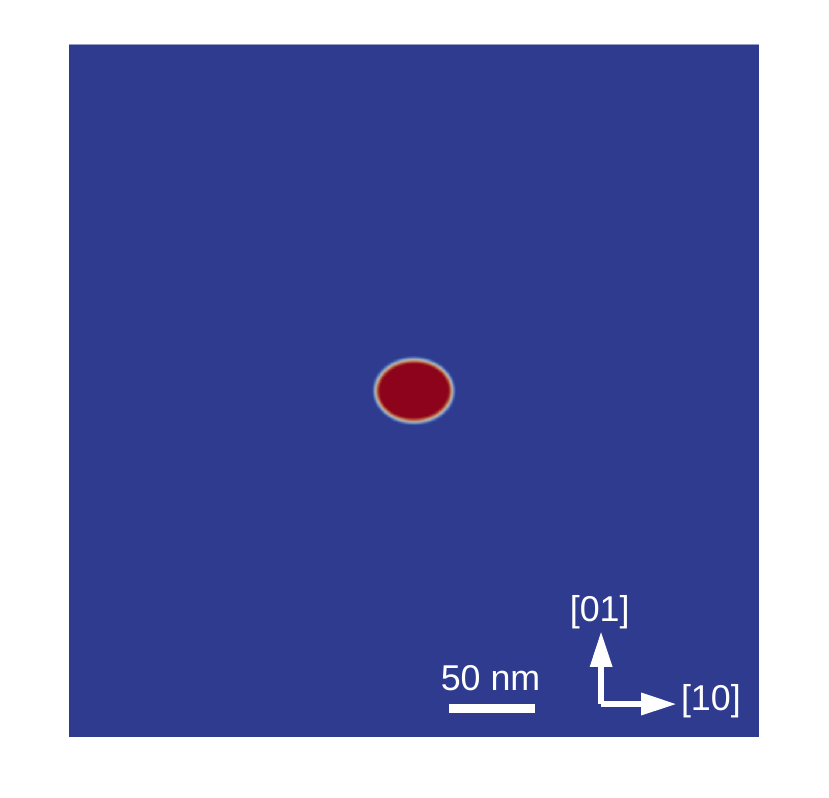}} 
\subfloat[\label{sfig:20_final}]{\includegraphics[scale=0.27]{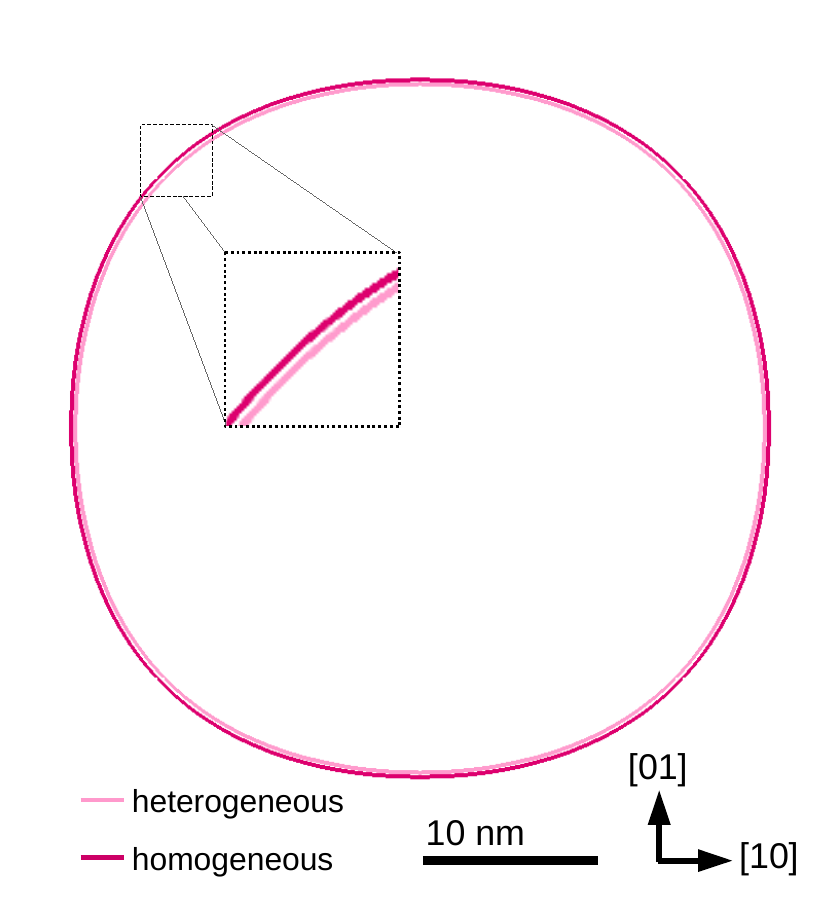}} 

\caption{(a) An example of the initial condition and computational domain size, shown for the precipitate with an initial area of $20^2\pi \textrm{ nm}^2$ and an elliptical shape.  (b) The final morphologies of the precipitates that have an initial area of $20^2\pi \textrm{ nm}^2$.  The dark pink line indicates the result when $C_{ijkl}^{matrix} = C_{ijkl}^{precip}$, and the light pink line indicates the result when $C_{ijkl}^{matrix} \neq C_{ijkl}^{precip}$.  For each case, the final shapes for the circle or ellipse initial condition are identical. The equilibrium morphology has four-fold symmetry and the choice of heterogeneous or homogeneous moduli does not significantly impact the precipitate morphology.  The inset illustrates more clearly that the results vary slightly depending on the choice of heterogeneous or homogeneous moduli.} \label{fig:small_equil_shape}

\end{figure}

\begin{figure}\centering
\subfloat[\label{sfig:20_het_energy}]{\includegraphics[scale=1]{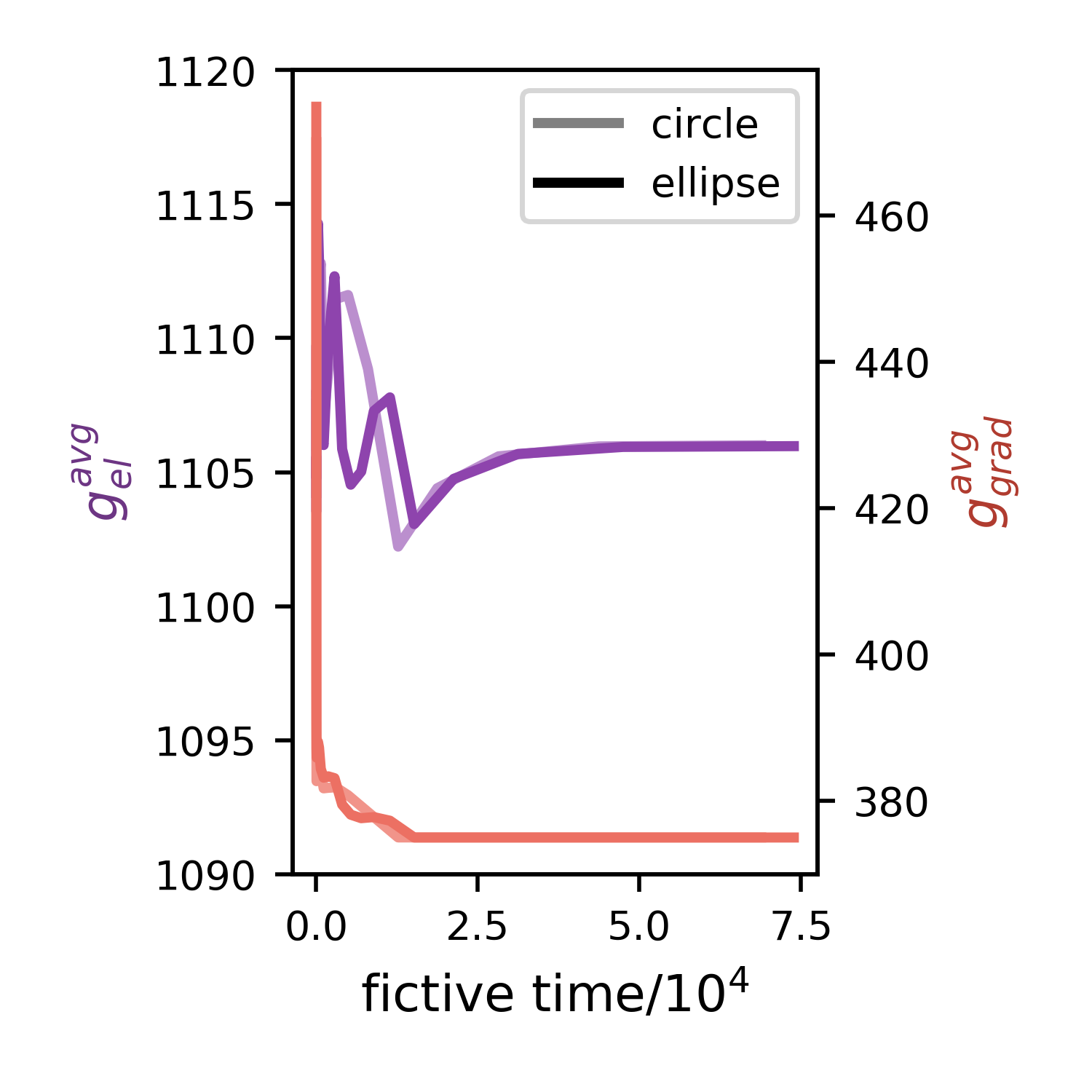}} 
\subfloat[\label{sfig:20_hom_energy}]{\includegraphics[scale=1]{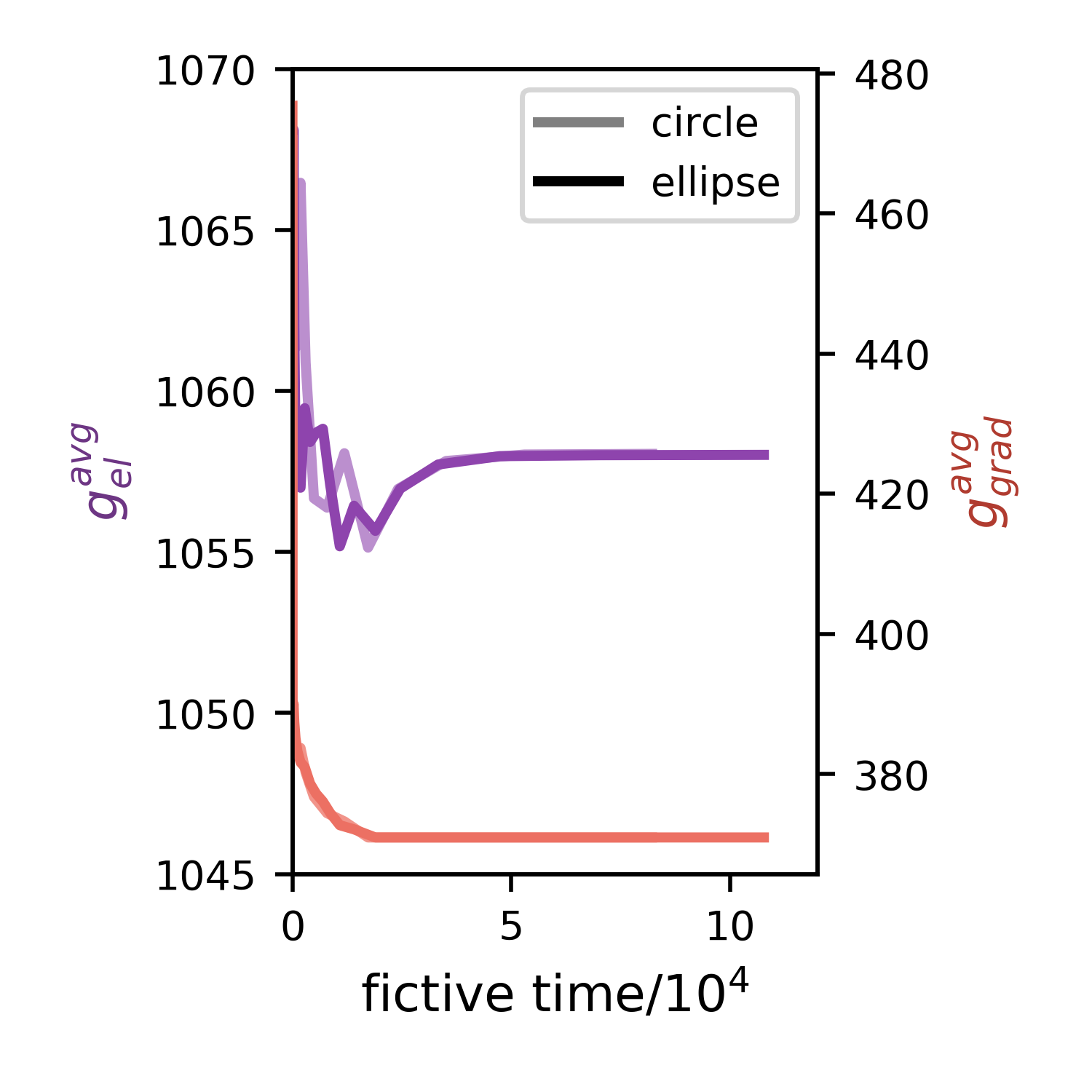}} 

\subfloat[\label{sfig:75_het_energy}]{\includegraphics[scale=1]{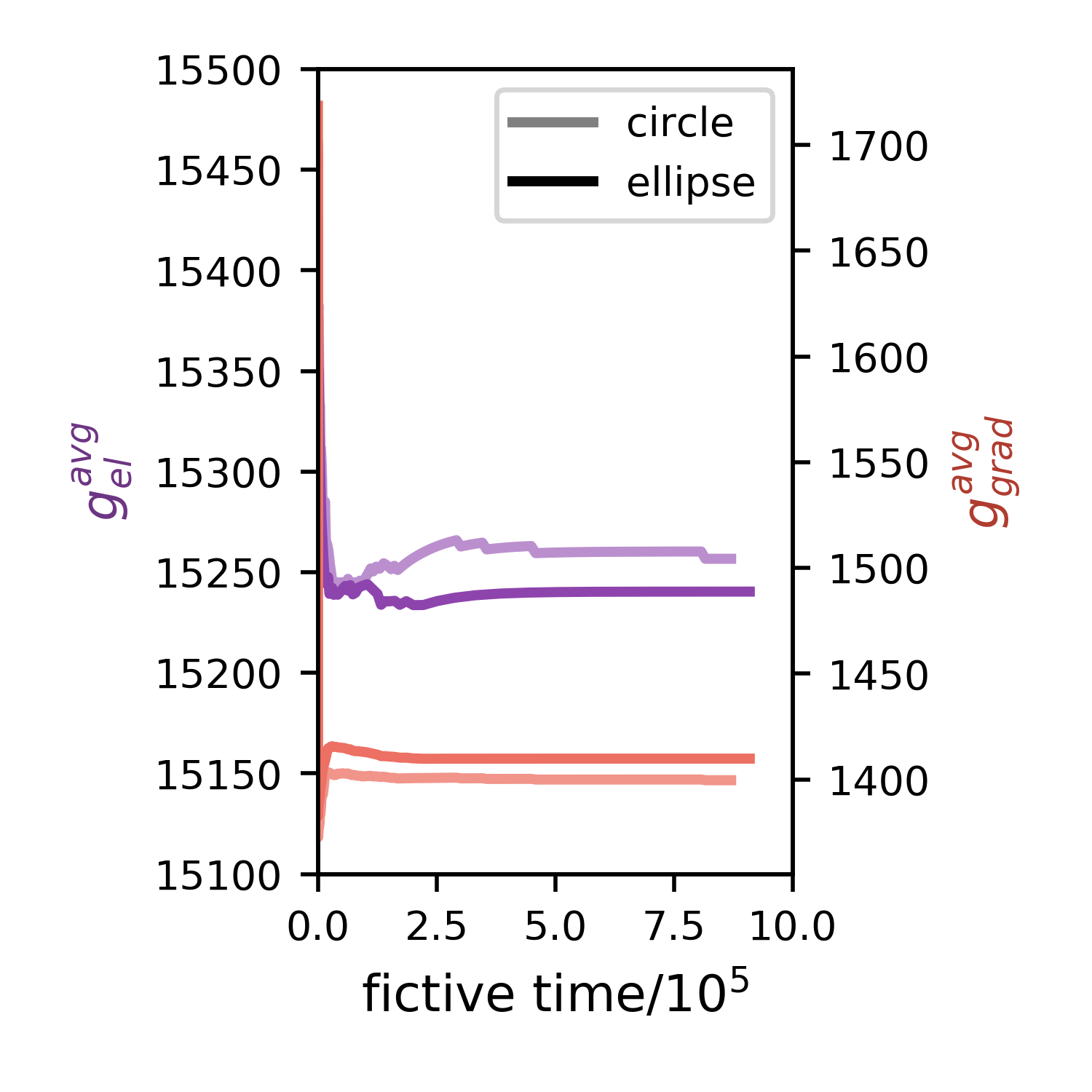}} 
\subfloat[\label{sfig:75_hom_energy}]{\includegraphics[scale=1]{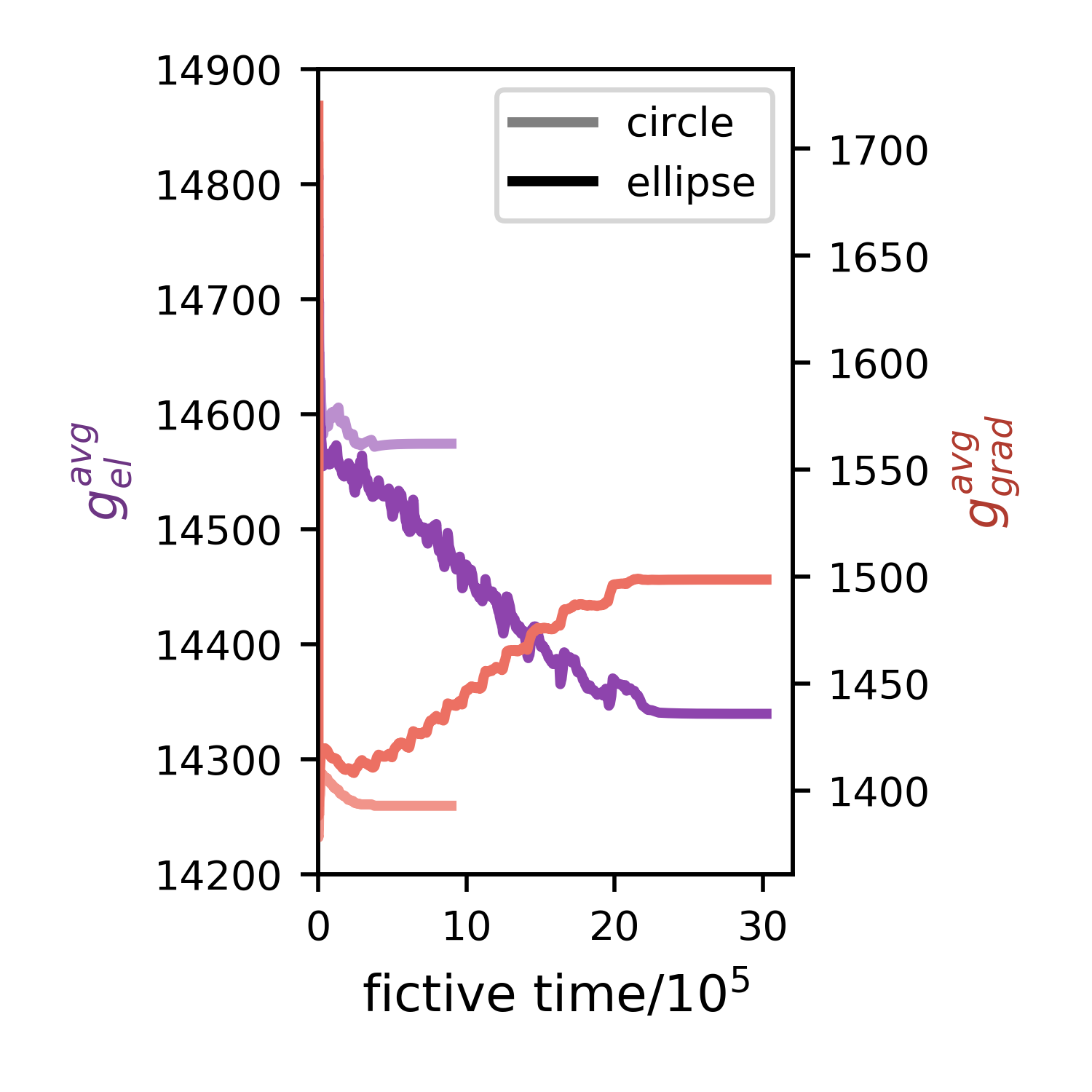}} 

\caption{The total elastic and gradient energies normalized by the precipitate area fraction, which evolves slightly.  (a), (b) initial precipitate area of $20^2\pi \textrm{ nm}^2$; (c), (d) initial precipitate area of $75^2\pi \textrm{ nm}^2$.  (a), (c)  $C_{ijkl}^{matrix} \neq C_{ijkl}^{precip}$; (b), (d) $C_{ijkl}^{matrix} = C_{ijkl}^{precip}$.  The darker lines indicate the circular initial condition, and the lighter lines indicate the elliptical initial condition; purple lines show the normalized elastic energy (left dependent-axis), and red lines show the normalized gradient energy (right dependent-axis).} \label{fig:elastic_energies}
\end{figure}

The equilibrium precipitate shapes are shown in Fig.~\ref{fig:small_equil_shape} for the precipitates with an initial area of $20^2\pi \textrm{ nm}^2$.  Regardless of the initial condition, the precipitates evolve to a shape with four-fold symmetry, indicating this is the energetically favored configuration. The $L'$ values for the precipitates are approximately 1, much smaller than the value at which bifurcation should occur (while the bifurcation point depends on the relationship between $C_{ijkl}$ values, it generally occurs for $L'>3$ \cite{thompson1994equilibrium}). At this particle size, the choice of the homogeneous modulus approximation does not significantly impact the precipitate shape versus that found with the heterogeneous moduli for the two phases.  In addition, the evolution of the total elastic and gradient energies normalized by the precipitate area fraction are shown in Fig.~\ref{fig:elastic_energies}, which we label as $g_{el}^{avg}$ and $g_{grad}^{avg}$; the gradient energy is closely related to the interfacial energy.  Figure~\ref{sfig:20_het_energy} shows the energies for the simulations in which the moduli are heterogeneous, and Fig.~\ref{sfig:20_hom_energy} shows the energies for the homogeneous moduli case. The final values of $g_{el}^{avg}$ and $g_{grad}^{avg}$ are the same for the simulations with initially circular and elliptical precipitates, as expected for their identical final morphologies.  In addition, the $g_{el}^{avg}$ is larger in the system with heterogeneous moduli, as expected for a stiffer precipitate.  The initial value of $g_{el}^{avg}$ for the elliptical initial shape is smaller than the final value, while the initial value of $g_{el}^{avg}$ for the circular initial shape is larger than the final value, indicating that the elastic energy contribution is reduced by the two-fold symmetry.  However, the total drop in gradient energy is larger in the systems for which the initial precipitates are elliptical and the magnitude of that drop is greater than the magnitude of the increase in elastic energy, indicating that the reduction in interfacial energy more than compensates for the increase in elastic energy, and thus, four-fold symmetry is preferred.

\begin{figure}\centering

\subfloat[\label{sfig:75_circ}]{\includegraphics[scale=0.3]{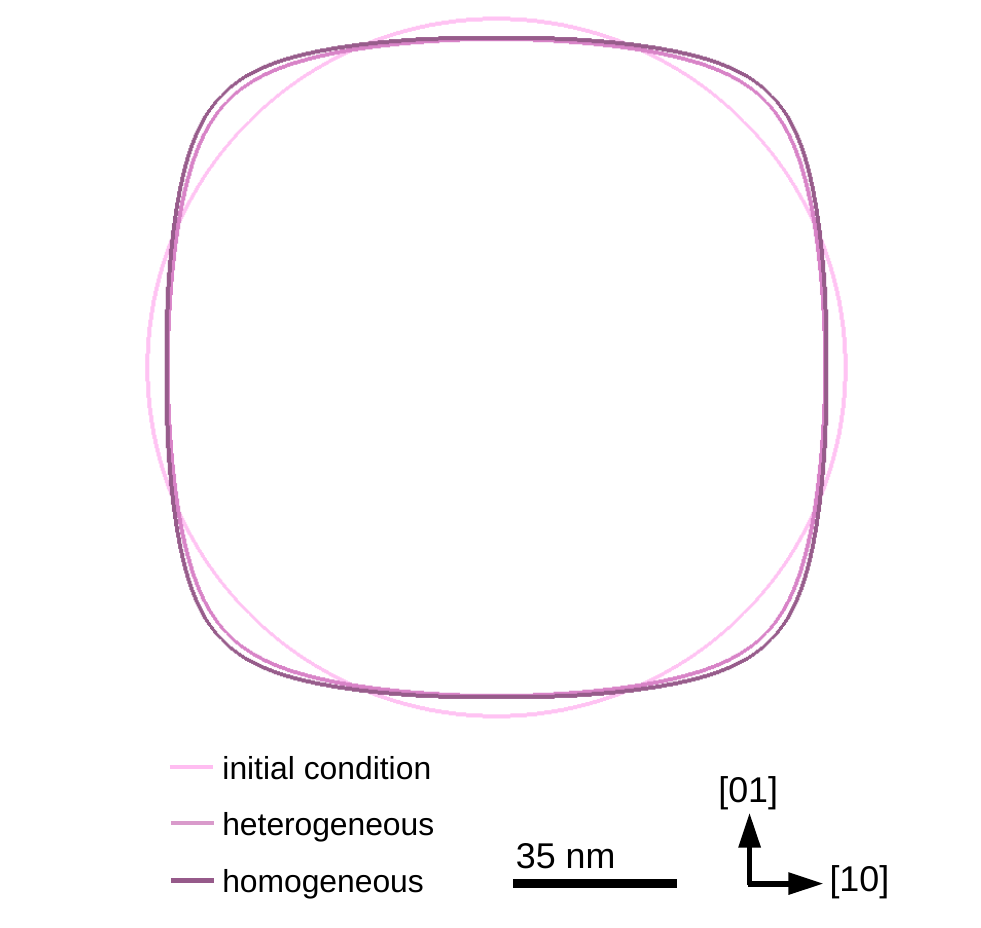}} 
\subfloat[\label{sfig:75_ell}]{\includegraphics[scale=0.3]{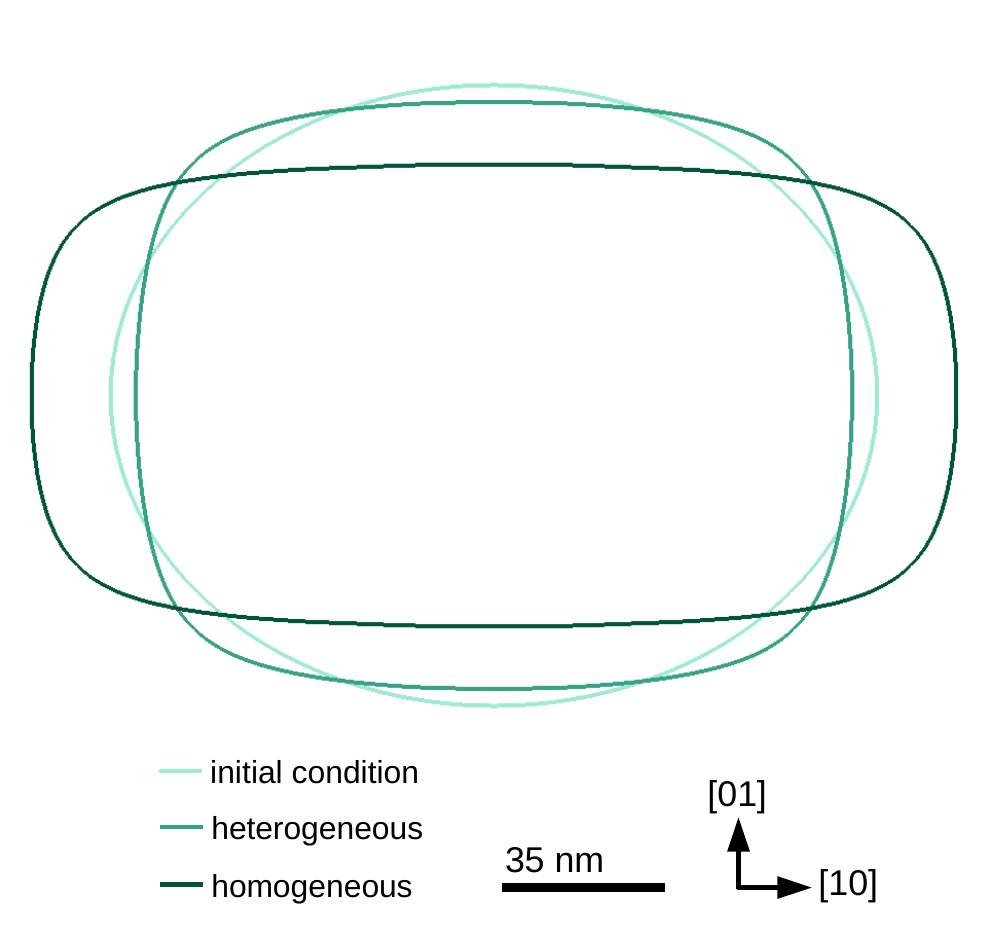}} 

\caption{The final morphologies of the precipitates that have an initial area of $75^2\pi \textrm{ nm}^2$.  (a) The final morphology when starting with an initially circular precipitate, and (b) the final morphology when starting with an initially elliptical precipitate.  The lightest, middle, and darkest shades indicate the initial morphology, the final morphology when $C_{ijkl}^{matrix} \neq C_{ijkl}^{precip}$, and the final morphology when $C_{ijkl}^{matrix} = C_{ijkl}^{precip}$, respectively. The equilibrium morphology has two-fold symmetry and the choice of heterogeneous or homogeneous moduli significantly impacts the precipitate morphology.} \label{fig:large_equil_shape}

\end{figure}

The equilibrium precipitate shapes are shown in Fig.~\ref{fig:large_equil_shape} for the precipitates with an initial area of $75^2\pi \textrm{ nm}^2$. Unlike the small particles, the large precipitates exhibit a significant variation in final shape.  First, the precipitates with the initially circular shape exhibit four-fold symmetry at equilibrium, while the precipitates with the initially elliptical shape exhibit two-fold symmetry. This indicates that the size of the precipitate has exceeded the bifurcation point \cite{thompson1994equilibrium}; the $L'$ values of these precipitates is 3.8 and 3.9 for the heterogeneous and homogeneous moduli systems, respectively.  The systems that start with a circular initial condition are unable to evolve toward the two-fold symmetry because there is no perturbation or anisotropy within the initial condition (or the numerical accuracy of the solver).  In addition, the precipitate shape varies significantly depending on whether the moduli for the phases are homogeneous or heterogeneous.  With even a 10\% increase in elastic stiffness of the precipitate, the equilibrium morphology is significantly more elongated. 

The energetics can provide further insight to the system behavior. Fig.~\ref{sfig:75_het_energy} shows the energies for the simulations in which the moduli are heterogeneous, and Fig.~\ref{sfig:75_hom_energy} shows the energies for the homogeneous moduli cases.  As with the smaller precipitates, the final value of $g_{el}^{avg}$ is larger for the heterogeneous moduli system than for the homogeneous moduli system.  In addition, the value of $g_{el}^{avg}$ is smaller for the precipitates with two-fold symmetry than those with four-fold symmetry, and while the value of $g_{grad}^{avg}$ is greater for the precipitates with two-fold symmetry, the increase in the total system interfacial energy is more than compensated for by the decrease in the total elastic energy.  The sawtooth pattern in the energy evolution appears to be related to the use of adaptive meshing and the discretized nature of the energy calculations and does not correlate with major morphological transitions.
  
This benchmark problem lends itself well to testing the effects of modifications to the model formulation and parameterization.  For example, a different interpolation function can be chosen for the elastic stiffnesses and misfit strain, which may impact properties across the interface.  In addition, the effect of interface width could be examined, as diffuse interface widths are often chosen as unphysically wide to reduce computational cost.  Furthermore, the problem can be used to test the effect of computational domain size on the result: long-range stresses relax over a significant distance, meaning that a domain size must be large enough to allow this relaxation, but too large of a domain can waste computational resources with no additional gain in accuracy. A final example is using symmetry to reduce the computational domain size; boundary conditions must be changed appropriately.

\section{Conclusion \label{sec:conclusion}}

In this work, we present our ongoing effort in developing numerical benchmark problems for phase field modeling with the proposal of two additional benchmark problems.  The problems, while essentially unrelated, target physics often incorporated into phase field models: one focuses on solidification and dendritic growth in a single-component system, while the other focuses on linear elasticity by simulating the equilibrium shape of an elastically constrained precipitate.  The benchmark problems involve simple formulations, reducing the need for time-consuming coding and debugging of complex free energies.  

The dendritic growth benchmark problem is used to demonstrate the effect of different numerical methods, especially time integration algorithms, on the simulated microstructure evolution. The problem is implemented twice: once using a bespoke code that employs finite differencing for spatial discretization and explicit Euler time integration without the use of mesh or time adaptivity, and once with a MOOSE-based application, which uses finite element spatial discretization, implicit time integration, and mesh adaptivity.  We present the final morphology, as well as the evolution of the free energy, solid area fraction, and tip velocity as simulation metrics.  We show that the free energy evolution is an unsuitable proxy for microstructure evolution, as relatively small differences in free energy may be the result of significantly different dendrite arm lengths; we suggest that this may be a concern for any model in which significant anisotropy exists. We also demonstrate that the choice of implicit time integrator can significantly change the microstructure evolution versus that simulated with the ``standard'' methods in the bespoke code.  This highlights the need for benchmark problems to test newly developed algorithms and implementations, especially with the current goal of incorporating quantitative phase field modeling results into ICME frameworks for materials design.

In contrast, the elastically constrained precipitate problem exploits the physical phenomenon of elastic instability and bifurcation to examine the effect of initial conditions and model approximations on the results.  In this problem, a precipitate is allowed to relax until no further shape change occurs.  Precipitates have initially circular and elliptical shapes, and the elastic moduli are either heterogeneous ($C_{ijkl}^{matrix}\neq C_{ijkl}^{precip}$), as is generally the case for real materials, or homogeneous ($C_{ijkl}^{matrix} = C_{ijkl}^{precip}$), a common approximation. We present final morphologies, dimension measurements, $L'$ values and normalized elastic and gradient energies as simulation metrics.  We show that the impact of initial condition and homogeneous modulus approximation is not significant at small precipitate sizes, but becomes important as the precipitate size increases. We also discuss how the benchmark problem may be adapted to test additional factors such as diffuse interface width and computational domain size.

In a broader sense, these benchmark problems will serve to keep phase field models and numerical algorithms and implementations in ``lockstep'': phase field modeling is no longer a niche method with a small core of highly experienced users, but one that has a rapidly expanding community that is constantly generating new ideas.  In particular, numerical solver frameworks have risen in popularity, partly due to the more complex coding required for parallel computing, which significantly lower the barrier to performing phase field simulations.  While not necessarily a feature in all framework solvers, frameworks often allow different algorithms to be used to perform the same task, e.g., time integration or mesh adaptivity.  However, not all numerical methods are appropriate to solve all physics problems; thus benchmark problems are a rapid means of evaluating the performance of a new numerical method versus other, known methods.  Further numerical benchmark problems are needed to model additional physics, such as fluid flow and electric fields, which are often included in more complex solidification models and electrochemistry, respectively.  We urge the community to continue to provide feedback for existing and possible additional benchmark problems, as well as to upload their results for comparison, by visiting the website, \url{https://pages.nist.gov/pfhub/}.

\section*{Acknowledgments}
This work was performed under financial assistance award 70NANB14H012 from U.S. Department of Commerce, National Institute of Standards and Technology as part of the Center for Hierarchical Material Design (CHiMaD). We gratefully acknowledge the computing resources provided on Blues and Fission, high-performance computing clusters operated by the Laboratory Computing Resource Center at Argonne National Laboratory and the High Performance Computing Center at Idaho  National Laboratory, respectively.  We thank Nana Ofori-Opoku for writing the bespoke dendritic growth code and Daniel Wheeler and Trevor Keller for their work on the website.  In addition, we thank all the participants of the CHiMaD phase field workshops held in Evanston, Illinois for their feedback.  A. M. J. particularly thanks Nana Ofori-Opoku and David Montiel for their many helpful discussions on dendritic growth.

\section*{Data availability statement}
The raw data and processed data required to reproduce these findings are available to download from \url{https://doi.org/10.18126/m2qs6z}. 

\bibliographystyle{model1-num-names}
\bibliography{benchmarkII.bib}

\end{document}